\documentclass[aps,superscriptaddress,pra,twocolumn,showpacs]{revtex4-1}
\usepackage{amsfonts, amsmath, amssymb, bm, bbm, graphicx, epsfig, epstopdf}
\usepackage{braket}
\usepackage{booktabs}
\usepackage{mathrsfs}
\usepackage{dcolumn}
\usepackage{times}
\usepackage{textcomp}
\usepackage{placeins}
\usepackage[
    colorlinks=true,
    linkcolor=teal,      
    citecolor=teal,      
    urlcolor=teal        
]{hyperref}
\usepackage[capitalize, nameinlink]{cleveref}

\usepackage{xcolor}
\usepackage{booktabs}
\usepackage{array}
\usepackage{tabularx}
\usepackage{makecell}
\usepackage{soul}
\definecolor{forestgreen}{RGB}{34,139,34}

\usepackage[normalem]{ulem} 
\usepackage{color}
\usepackage{tikz}
\usepackage{subcaption}
\usepackage{graphicx}
\usetikzlibrary{positioning, fit, arrows.meta, backgrounds, shadows, shapes.geometric, calc}
\newcounter{panel}
\renewcommand{\thepanel}{\alph{panel}}
\newcommand{\panellabel}[1]{%
    \refstepcounter{panel}%
    \textbf{(\thepanel)}\label{#1}%
}

\usepackage{ragged2e}

\makeatletter
\long\def\@makecaption#1#2{%
  \par\vskip\abovecaptionskip
  \begingroup
    \small
    \setlength{\parindent}{0pt}
    \noindent\justifying #1: #2\par
  \endgroup
  \vskip\belowcaptionskip
}
\makeatother

\usetikzlibrary{decorations.pathmorphing}
\definecolor{photonblue}{RGB}{40, 100, 200}
\definecolor{decayred}{RGB}{160, 30, 70}

\newcommand{\kket}[1]{|#1\rangle\!\rangle}
\newcommand{\bbra}[1]{\langle\!\langle #1|}

\begin{abstract}
We develop an analytic perturbative framework that enables the analysis of small Markovian errors in probabilistic, photon-heralded quantum operations between non-interacting emitters.
Building on and extending the Zero-Photon-Generation (ZPG) framework, we derive closed-form perturbative solutions that capture both ideal (zero-order) and noisy (low-order) gate dynamics conditioned on time-integrated photon counting. Our framework provides analytic solutions to process matrices and Pauli error weights up to leading order, bridging the gap between detailed physical imperfections of a system and its corresponding abstract Pauli noise models. 
Moreover, our framework captures imperfections across the full physical system stack, from photon generation to optical manipulation. We benchmark the resulting perturbative predictions on a repeat-until-success $\mathsf{CZ}$ gate against numerical simulations, demonstrating accurate modeling of source-induced noise, and then apply the same framework to analyze coherent phase-shifter miscalibrations as a representative example of optical-manipulation errors.
 The methods developed in this work enable physics-informed parameter tuning to optimize gate designs and develop tailored quantum error correction protocols toward fault-tolerant quantum computing using hybrid light--matter quantum systems.
\end{abstract}

\begin{document}
\title{Characterization of errors in photon-heralded quantum operations\\between non-interacting quantum emitters
}

\author{Mahsa Karimi}
\email{mahsa.karimi1@ucalgary.ca}
\affiliation{Institute for Quantum Science and Technology, and Department of Physics \& Astronomy, University of Calgary, 2500 University Drive NW, Calgary, Alberta T2N 1N4, Canada}

\author{Samuel Mister}
\affiliation{Quandela, 7 Léonard De Vinci, 91300 Massy, France}

\author{Christoph Simon}
\affiliation{Institute for Quantum Science and Technology, and Department of Physics \& Astronomy, University of Calgary, 2500 University Drive NW, Calgary, Alberta T2N 1N4, Canada}

\author{Stephen C. Wein
}
\affiliation{Quandela, 7 Léonard De Vinci, 91300 Massy, France}

\maketitle

\section{Introduction}
Quantum computing offers powerful advantages for simulating quantum many-body systems and for accelerating specific computational tasks beyond classical capabilities, driving major advances in hardware, control, and error-correction toward scalable and fault-tolerant platforms \cite{feynman2018simulating,shor1999polynomial,grover1996fast,preskill2018quantum,bluvstein2024logical}. A promising architectural approach exploits light–matter interfaces, where coherently emitted single photons from stationary spin qubits are interfered in linear-optical networks and measured with photon-number-resolving detectors. This hybrid approach enables heralded entanglement generation between otherwise non-interacting qubits and underlies several proposals for universal quantum computation and simulation \cite{barrett2005efficient,lim2006repeat,de2024spin,uysal2025spin,ruskuc2025multiplexed}. Realizing a utility-scale quantum computer requires high-fidelity operation, and hence, the study of hardware errors is of paramount importance. In particular, some of the most disruptive quantum algorithms in cryptography and chemistry may require upward of a trillion operations \cite{shor1999polynomial, reiher2017elucidating}, which is expected to require quantum error correction with physical error rates below threshold \cite{aharonov1997fault}. Such studies become more challenging in the measurement-driven light--matter architectures, owing to the probabilistic nature of their elementary building blocks and the large Hilbert space of the flying qubits, which arises both from the infinite-dimensional Fock basis and from their time--frequency degrees of freedom.

\begin{figure*}[!t]
\centering

\begin{subfigure}[c]{0.59\textwidth}
    \centering
    \includegraphics[width=\linewidth]{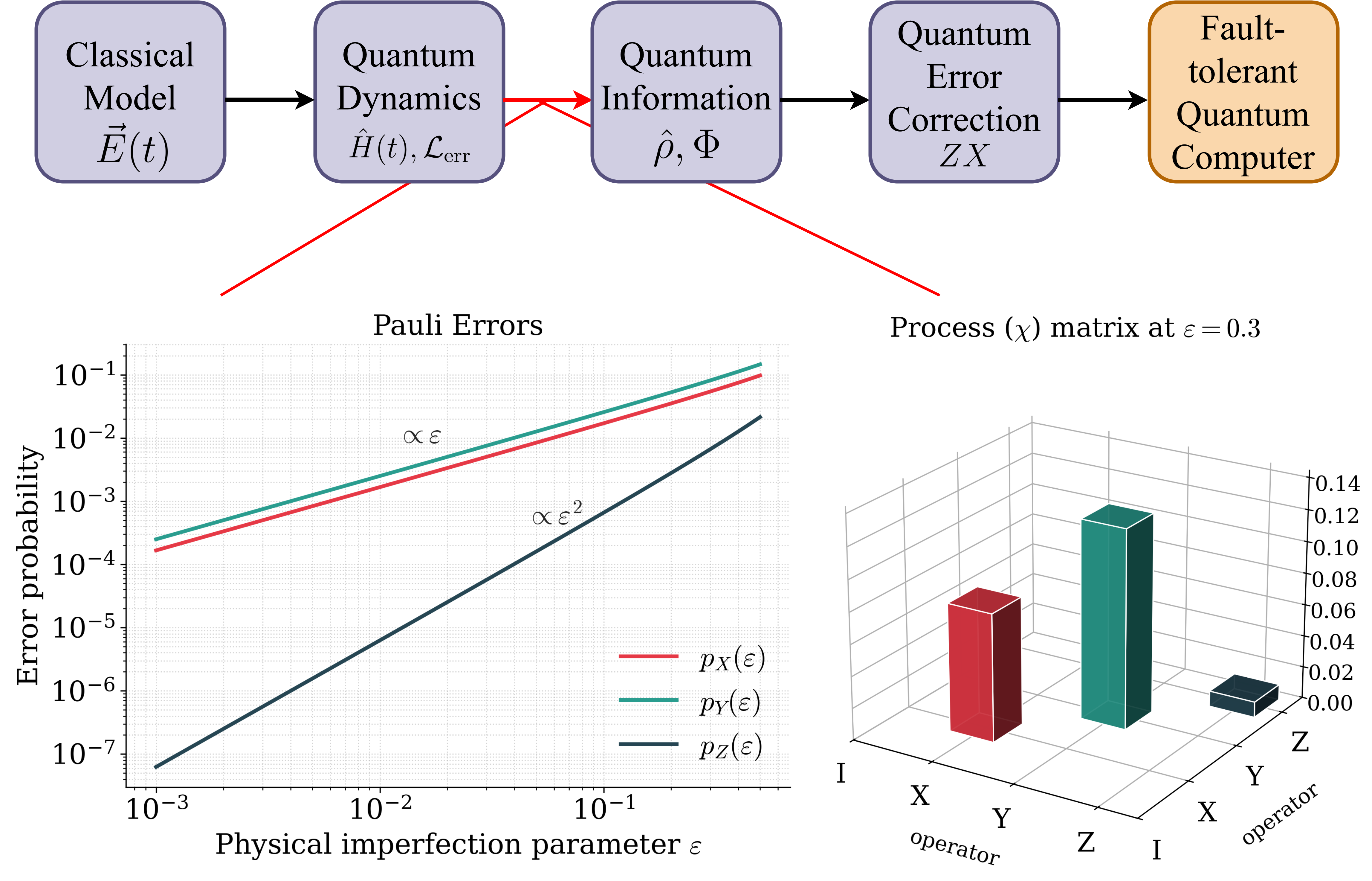}
    \caption{}
    \label{subfig:example}
\end{subfigure}
\begin{subfigure}[c]{0.4\textwidth}
\centering
\begin{tikzpicture}[
    font=\small,
    node distance=7mm and 7mm,
    box/.style={
        draw,
        rounded corners=2mm,
        align=center,
        inner ysep=6pt,
        inner xsep=7pt,
        line width=0.6pt
    },
    proc/.style={box, fill=blue!4, draw=blue!55!black},
    emph/.style={box, fill=orange!8, draw=orange!70!black, line width=0.8pt},
    side/.style={box, fill=gray!10, draw=gray!60!black, font=\footnotesize},
    arr/.style={-Stealth, line width=0.8pt, draw=gray!80!black},
    mainarr/.style={-Stealth, line width=0.9pt, draw=blue!70!black},
    workarr/.style={-Stealth, line width=1.0pt, draw=orange!80!black},
]

\node[proc, text width=5.1cm] (mid) {
  \textbf{Initial framework (ZPG)}
\cite{wein2020analyzing}
};

\node[proc, text width=5.1cm, above=7mm of mid] (num) {
  \textbf{Numerical tool}\\[-2pt]
  \texttt{ZPGenerator }\cite{wein2024simulating}
};

\draw[mainarr, dashed] (mid) -- (num);

\coordinate (row) at ([yshift=-20mm]mid.south); 

\node[side, text width=2.55cm, anchor=east] (yb) at ([xshift=-3mm]row) {
  \textbf{Deterministic-gate}\\[-2pt]
  \textbf{perturbation}~\cite{karimi2026comparing}
};

\node[emph, text width=2.55cm, anchor=west] (pert) at ([xshift=+3mm]row) {
  \textbf{General}\\[-2pt]
  \textbf{perturbative}\\[-2pt]
  \textbf{expansion}\\
  \text{[This work]}
};

\draw[workarr, shorten <=2pt, shorten >=2pt]
  (mid.south) -- node[
    midway, fill=white, inner sep=1.5pt,
    text=orange!80!black, font=\footnotesize, rounded corners=1pt
  ] {Apply perturbation} (pert.north);

\draw[workarr, shorten <=2pt, shorten >=2pt] (yb.east) -- (pert.west);

\end{tikzpicture}
\caption{}
\label{subfig:diagram}
\end{subfigure}

\caption{
(a)
Conceptual pipeline from physical models to fault-tolerant quantum-computing design. Device-level descriptions are mapped to open-system quantum dynamics, which are then reduced to quantum-information metrics such as process matrices or stochastic Pauli error models. These qubit-level descriptions provide input for quantum-error-correction analysis and ultimately fault-tolerant logic. This work targets the highlighted step for probabilistic light--matter architectures by mapping hardware imperfections to leading-order analytic process matrices.
(b)
Relation to prior approaches. The Zero-Photon Generator (ZPG) framework gives analytic solutions under restricted error models \cite{wein2020analyzing}, while the ZPGenerator package provides exact numerical solutions for more general imperfections \cite{wein2024simulating}. Perturbative methods have also been used for deterministic gates generated by direct deterministic interactions \cite{karimi2026comparing}. This work combines perturbation theory with ZPG to obtain analytic low-order error models for probabilistic photon-heralded operations between non-interacting qubits.}
\label{fig:framework-overview}
\end{figure*}
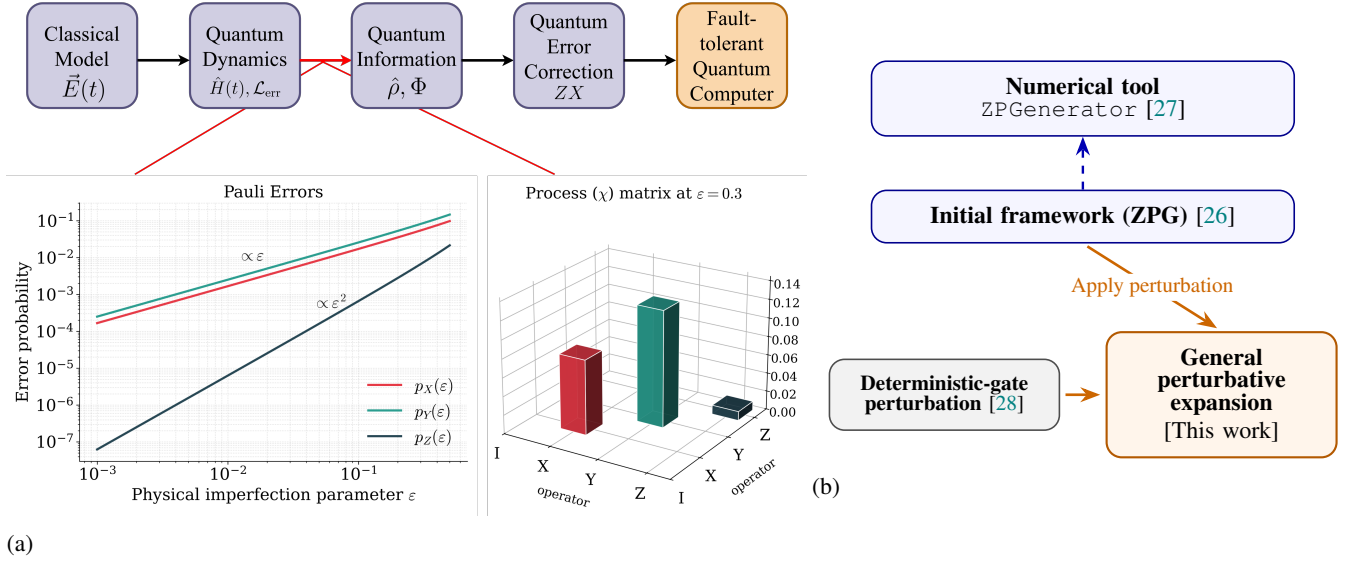

To analyze errors arising from a quantum system’s interaction with its environment, one often begins with a physical description of the underlying hardware and derives an effective Lindbladian that captures the resulting open-system quantum dynamics. For instance, the evolution of a quantum emitter can be derived by quantizing a classical model of the surrounding electromagnetic environment, with the parameters entering the resulting open-system description often informed by classical electromagnetic simulations \cite{gustin2025dissipation,franke2019quantization}.
This hardware-level description enables the design of physical qubits and gates, but assessing scalability requires translating these dynamics into a stochastic qubit-level noise channel, commonly represented by a process ($\chi$) matrix or error channels \cite{de2024spin}. In particular, the diagonal elements of the $\chi$ matrix in the Pauli basis correspond to effective stochastic Pauli error rates, which serve as a primary input for selecting quantum error-correcting codes and estimating logical error rates \cite{geller2013efficient,wallman2016noise,dankert2009exact,katabarwa2015logical,takou2025estimating}. These logical error rates determine whether the device operates below the fault-tolerance threshold, a prerequisite for scalable quantum computation under the threshold theorems \cite{shor1995scheme,steane1996multiple,aharonov1997fault,knill1998resilient,kitaev2003fault,fowler2012surface,gidney2021stim}.
This hierarchy is shown in \cref{fig:framework-overview}a.
In this work, we focus on the critical interface between microscopic hardware imperfections and their emergent qubit-level noise models (highlighted part in \cref{fig:framework-overview}a), enabling threshold assessment and principled design of scalable, high-fidelity architectures based on hybrid light--matter quantum systems.

In analyzing errors in light–matter interactions within measurement-driven architectures, analytical descriptions are particularly valuable because they provide insight into the high-dimensional error landscape and the combined effect of different errors. 
In this context, the Zero-Photon Generation (ZPG) framework was introduced to describe photon-emission processes by defining a zero-photon–conditioned generator for the no-emission evolution and recursively deriving the corresponding multi-photon processes \cite{carmichael1993open,wein2020analyzing,wein2021modelling}.
This framework enabled analytic fidelity estimates under simplifying assumptions. However, incorporating realistic imperfections may render the analytic expressions intractable, limiting their utility for rapid design iteration and parameter optimization.
Moreover, the existing applications have focused on fixed-input-state fidelities rather than gate-level error channels and average fidelities, which are required for composable circuit-level noise modeling in quantum-computing design. While numerical simulations can evaluate state fidelities and error channels under broad error models \cite{wein2024simulating}, they generally provide less direct structural insight into noise biases and multi-parameter trends compared to analytical approaches.

Perturbative methods are particularly well-suited for modeling fault-tolerant quantum-computing architectures, as quantum error correction targets regimes with small physical imperfections. The idea of treating imperfections as controlled perturbations has been used before to estimate state and gate fidelities in deterministic gate schemes mediated by direct interactions \cite{karimi2026comparing}, whereas the ZPG framework addresses probabilistic, measurement-driven light–matter architectures. In this work, we combine these two ideas to develop a perturbative framework for photon-heralded light–matter operations that retains analytical tractability in the low-noise regime.

The resulting framework applies to conditional spin maps induced by photon counting, including unitary gates, projections, and more general non-unitary operations. It yields leading-order analytic expressions for state fidelities, gate fidelities, and process ($\chi$) matrices, thereby providing an interpretable route from microscopic error mechanisms to qubit-level noise models. \Cref{fig:framework-overview}b summarizes how this work extends prior ZPG-based analytic methods, numerical ZPG simulations, and perturbative treatments of deterministic gates.

The paper is structured as follows. \cref{sec:method} introduces the mathematical framework, beginning with a review of the Zero-Photon-Generation (ZPG) framework \cite{carmichael1993open,wein2020analyzing,wein2021modelling, wein2024simulating} in \cref{sec:ZPG}. In \cref{sec:zero-order,sec:first-order}, we derive expressions for the zero and the first-order solutions. 
In \cref{sec:fidelity}, we show how these perturbative expansions to provide gate errors and effective noise models. In \cref{sec:application}, we apply the method to the repeat-until-success (RUS) $\mathsf{CZ}$ gate \cite{lim2005repeat, lim2006repeat}, deriving the associated Pauli error expressions, and benchmarking them against exact simulations obtained with the ZPGenerator package. We conclude and provide an outlook in \cref{outlook}.\\

\section{Method}\label{sec:method}

In this section, we introduce an analytical framework for evaluating error channels in photon-mediated conditional operations. In what follows, we use $\mathbb N :=\{0,1,\ldots\}$ to denote the non-negative integers, and $N$ to denote the number of photon-number-resolving detectors (PNRDs). We use a calligraphic font for superoperators such as the Lindbladian $\mathcal L$, and we use bold symbols for vectors of integers, e.g., $\mathbf m\in\mathbb N^N$. We use $\mathcal I$ for the identity channel, and $\mathbb I$ for the identity matrix. For two vectors of integers $\mathbf n,\mathbf m \in \mathbb N^N$, we say $\mathbf n\leq \mathbf m$ if $n_i\leq m_i$ for all $i\in \{1,\ldots, N\}$. For a given perturbation parameter, such as $\varepsilon$, we use the notation
\begin{align}
\lambda = \lambda^{(0)} + \varepsilon \lambda^{(1)} + \varepsilon^2 \lambda^{(2)} + \cdots,
\end{align}
for a perturbative expansion of a quantity $\lambda$. That is, we use $\lambda^{(i)}$ as the $i$-th order perturbation to $\lambda$. 
Lastly, we use the notation $\kket{M}$ and $\bbra{M}$ to denote the column vectorization of a matrix $M$ and its conjugate-transposed row vectorization, respectively.

\subsection{Photon-counted light--matter interaction}\label{sec:ZPG}

Let us model the light--matter interaction beginning with a Lindblad master equation
\begin{align}
\dot\rho = \mathcal L[\rho],
\end{align}
where $\rho$ is the density operator of the matter system, and does not include the states of photons.
We then consider $M$ quantum emitters that each emit one photon into at least one of $N$ modes where $N\ge M$. We allow the possibility for one or more emitters to emit photons into multiple modes, such as polarization modes. The photons are then interfered using an arbitrary linear-optical transformation represented by the unitary matrix $U$ and measured with number-resolving detectors, post-selecting on a particular photon pattern $\mathbf m \in \mathbb N^N$. We assume an initial state $\rho_{\mathrm{init}}$ and let us denote the state obtained by measuring photon pattern $\mathbf m$ as $\rho_{\mathbf m}$.
Also, let $\mathcal P_{\mathbf m}$ denote the effective map that describes the evolution conditioned on the pattern $\mathbf m$. That is:
\begin{align}
\rho_{\mathbf m} = \mathcal P_{\mathbf m}(t,t_0)[\rho_{\mathrm{init}}],
\end{align}
for time-integrated detection of photons during the time interval $(t_0,t)$.

In what follows, we assume detectors are perfect, but we highlight that detector inefficiencies can be taken into account in a straightforward way, as detailed in \cref{sec:eta}. Therefore, our approach can be used to capture detector inefficiencies in a non-perturbative manner, and we can study realistic scenarios where $\eta$ is far from $1$.

Following the ZPG framework \cite{wein2020analyzing, wein2024simulating}, we can compute $\mathcal P_0$ (i.e., the evolution conditioned on measuring no photons) by solving the ZPG master equation
\begin{align}
\dot \rho = \mathcal L_{\mathrm{ZPG}}[\rho],
\end{align}
defined by
\begin{align}
\label{eq:zpg}
\mathcal L_{\mathrm{ZPG}} = \mathcal L - \sum_{i} \Gamma_i\mathcal S_i,
\end{align}
where $\mathcal S_i$ corresponds to the superoperator defining the emitter transition associated with the emission of a photon into the $i$-th mode \cite{carmichael1993open} at rate $\Gamma_i$. Here, $\mathcal S_i[\bullet] = \sigma_i \bullet \sigma_i^\dagger$ where $
\sigma_i$ is the lowering operator of the emitter transition producing photons in the $i$-th emission mode. Following the assumptions of the ZPG framework, these operators are assumed to be proportional to the electromagnetic field mode operators as a consequence of the input-output theory \cite{gardiner1985input} with vacuum input states, thus each system transition creates a photon in its corresponding mode (see \cref{app:proofs} for more details). We can exponentiate $\mathcal L_{\mathrm{ZPG}}$ to obtain the zero-photon conditional map
\begin{align}
\mathcal P_{0} = \mathscr{T} \exp\left( \int_{t_0}^t \mathcal L_{\mathrm{ZPG}}(t')\, \mathrm dt' \right),
\end{align}
where $\mathscr{T}$ denotes the time-ordering operator. Let the linear-optical transformation be represented by the linear-optical mode transformation matrix $U$ and define
\begin{align}\label{eq:def-of-D}
\mathbf{D} = U\boldsymbol{\sigma},
\end{align}
for $\boldsymbol{\sigma} = (\sigma_i)_i$ being the vector of lowering operators. We can interpret the operators $\mathbf D$ as the lowering operators evolved under the interferometer. We also define the jump superoperators $\mathcal J_i[\bullet]:= D_i \bullet D_i^\dagger$, that can be interpreted as the action on the emitter system when detecting one photon in the output mode $i$.
Then, via the following recursive relation \cite{wein2020analyzing}, we can compute arbitrary patterns
\begin{align}\label{eq:ZPG}
\mathcal P_{\mathbf m}(t,t_0) = \sum_{j=1}^N \int_{t_0}^t \mathcal P_0(t,t') \, \mathcal J_j \, \mathcal P_{\mathbf m - \mathbf e_j} (t',t_0)\, \mathrm dt',
\end{align}
where we use $\mathbf e_j$ to denote the $j$-th standard basis element, and we set any $\mathcal P_{\mathbf m - \mathbf e_j}$ to zero if $\mathbf m - \mathbf e_j$ has a negative entry.

\begin{figure}
    \centering
    \includegraphics[width=1\linewidth]{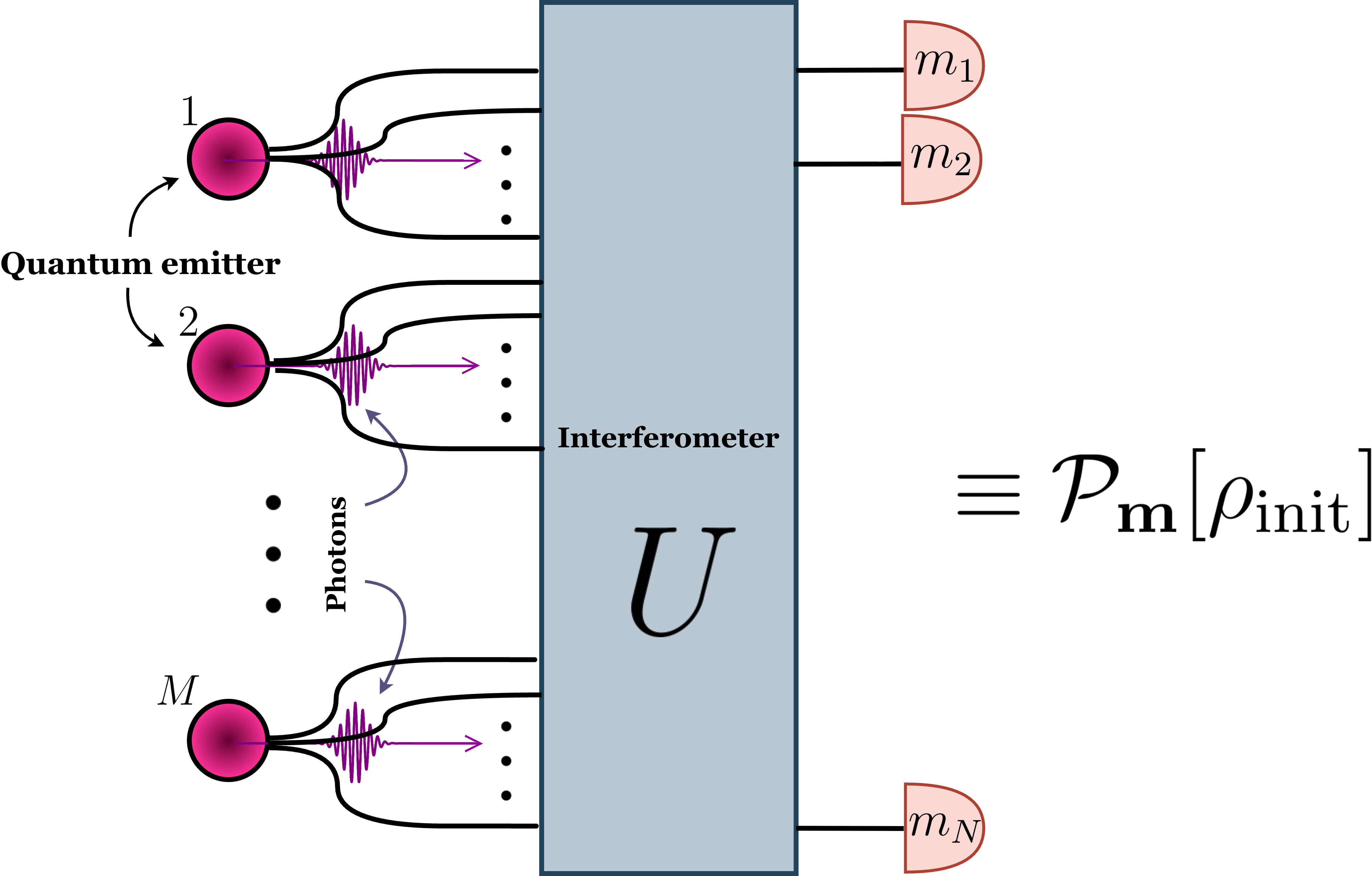}
    \caption{Illustration of the general scenario captured by the ZPG framework \cite{wein2024simulating}. The framework considers $M$ emitters where each emitter can emit a photon into $N\ge M$ different modes. The emitted photons interfere and are then measured with photon-number resolving detectors (PNRDs) producing a pattern $\mathbf m\in\mathbb N^N$. The initial state $\rho_\mathrm{init}$ of the ensemble of emitters collapses into a state proportional to $\mathcal P_{\mathbf m}[\rho_{\mathrm{init}}]$. The probability of obtaining measurement pattern $\mathbf m$ is $\mathrm{tr}(\mathcal P_{\mathbf m}[\rho_{\mathrm{init}}])$.}
    \label{fig:placeholder}
\end{figure}

\subsection{Zero-order evolutions}\label{sec:zero-order}
To extend the above framework to small perturbations of the ideal dynamics that represent error mechanisms, we begin by defining a noisy Lindbladian
\begin{align}\label{eq:lindblad-expansion}
\mathcal L = \mathcal L^{\mathrm{ideal}} + \sum_{k} \gamma_k \mathcal L_k^{\mathrm{err}},
\end{align}
where $\mathcal L_k^{\mathrm{err}}$ corresponds to the $k$-th error term. These generators may describe coherent Hamiltonian errors or dissipative error processes. Moreover, we assume that the photon emission dynamics are described by:
\begin{equation}
    \mathcal{L}^\mathrm{ideal}[\bullet] = -i[H, \bullet] + \sum_i \Gamma_i \mathcal{D}(\sigma_i)[\bullet],
\end{equation}
where
\begin{align}
\label{eq:dissipation}
\mathcal D(A)[\bullet] = A\bullet A^\dag - \frac12 \{A^\dag A, \bullet\}.
\end{align}
Unlike standard perturbative treatments in which the zero-order evolution is often unitary, the ideal dynamics considered here explicitly rely on dissipative mechanisms, such as spontaneous emission. Accordingly, $\mathcal{L}^{\mathrm{ideal}}$ includes the desired decay terms that produce the photons used by the protocol.

Although the perturbative framework can be applied to general systems, in many cases it is not possible to obtain tidy analytic solutions. To facilitate the derivation of insightful expressions, we focus on a simplified but relevant situation where, in the absence of errors, the emitter Hamiltonian $H$ is time independent and can be analytically diagonalized. This is reasonable given that analytic perturbative methods are generally only useful if the zero-order is analytically solvable. It is often the case for protocols relying on near-instantaneous excitation followed by time-independent spontaneous emission. We furthermore assume that the operators $\sigma_i$ are lowering operators of energy eigenstates, so that $\sigma_i^\dagger\sigma_i$ is diagonal in the energy eigenbasis, and that all $\sigma_i$ corresponding to the different transitions of the same emitter commute. This latter assumption holds, for example, in the case where the emission of a photon brings the emitter into a ground state. These assumptions are expanded upon in \cref{app:proofs}.

The zero-order ZPG $\mathcal L_{\mathrm{ZPG}}^{\mathrm{ideal}}=\mathcal L^{\mathrm{ideal}} - \sum_{i} \Gamma_i\mathcal S_i$ can be rewritten as an effective non-Hermitian Hamiltonian \cite{fischer2018scattering} by considering that the jump term $\Gamma_i\mathcal{S}_i$ will cancel out the first term of \cref{eq:dissipation}, leaving a clean way to combine the amplitude damping term $(\Gamma_i/2)\{\sigma_i^\dag\sigma_i,\bullet\}$ with the unitary dynamics,
\begin{align}
\mathcal L_{\mathrm{ZPG}}^{\mathrm{ideal}}[\bullet] = -i(H_{\mathrm{eff}} \bullet - \bullet H_{\mathrm{eff}}^\dagger),
\end{align}
where $H_{\mathrm{eff}}:= H - \frac{i}2 \sum_{i} \Gamma_i\sigma_i^\dag\sigma_i$ for $i$-th emission mode decay rate $\Gamma_i$. As a result, the zero-order zero-photon conditional map is
\begin{align}
\mathcal P_0^{(0)}(t,t_0)[\bullet] = K(t,t_0) \bullet K^\dagger(t,t_0),
\end{align}
where $K(t,t_0) = \mathscr{T}\exp\left(-i \int_{t_0}^t H_{\mathrm{eff}}(t') \mathrm dt'\right)$ in the general case, which simplifies to

\begin{align}
\mathcal P^{(0)}_0(t,t_0)[\bullet] = e^{-iH_{\mathrm{eff}}(t-t_0)} \bullet e^{iH_{\mathrm{eff}}^\dag(t-t_0)},
\end{align}
for time-independent evolution.

Our first result is to find a closed-form solution for the ideal evolution (zero-order ZPG), but conditioned on arbitrary photon patterns, i.e., $\mathcal P_{\mathbf m}^{(0)}$. We then find that
\begin{align}\label{eq:zero-order-closed-form}
\begin{split}
&\mathcal P^{(0)}_{\mathbf m}(t,t_0)[\bullet] = \\
&
f_{\mathbf m}(t-t_0)\cdot e^{-iH_{\mathrm{eff}}(t-t_0)} \mathbf D^{\mathbf m} \bullet \mathbf D^{\mathbf m}{}^\dagger e^{iH_{\mathrm{eff}}^\dagger(t-t_0)},
\end{split}
\end{align}
where $\mathbf D^{\mathbf m} := D_1^{m_1} \cdots D_N^{m_N}$. This notation is well-defined, since our previous assumption that all $\sigma_i$ commute implies that all $D_i$ also commute. When the decay rates for all emitters are equal, $\Gamma_i=\gamma$, the functions $f_{\mathbf m}$ are simply
\begin{align}\label{eq:f-solution}
\begin{split}
f_{\mathbf m}(t) = \frac{1}{\mathbf m!} (1-e^{-\gamma t})^{|\mathbf m|}.
\end{split}
\end{align}
Here, we use $|\mathbf m|=\sum_{j=1}^N m_j$, and $\mathbf m!=\prod_{j=1}^N m_j!$.  We provide the derivation of \cref{eq:zero-order-closed-form} in \cref{app:proofs}. 

The functions $f_{\mathbf m}$ have useful physical interpretations as the time-ordered integrals over possible emission times. To see this, consider the single-mode case $N=1$ where
$
f_m(t) = \frac{1}{m!} (1-e^{-\gamma t})^m
$
is proportional to a Poisson distribution over $m$ with a mean of $1-e^{-\gamma t}$. In other words, $f_m$ captures the unnormalized statistical weight associated with $m$ emission events.
Similarly, the multi-mode factor $f_{\mathbf m}$ captures the statistical weight for a photon-number pattern $\mathbf m \in \mathbb N^N$ arising from $N$ independent Poisson distributions. This classical unnormalized statistical factor is then modulated by the quantum dynamics through $H_\mathrm{eff}$ and weighted by the multi-mode transition amplitudes encoded in $\mathbf{D}^\mathbf{m}$.

Note that \cref{eq:zero-order-closed-form} implies that pure input states remain pure under the conditional map $\mathcal P_\mathbf{m}^{(0)}$ after normalization, since the map has a single Kraus operator. This is intuitive because, in the ideal case, all dissipative channels are perfectly monitored by photon detectors. Nevertheless, it is not immediately obvious from the general ZPG framework, which is formulated at the level of density matrices.

\subsection{First-order evolutions}\label{sec:first-order}

Now, we consider the imperfections introduced by $\mathcal L_k^\mathrm{err}$ terms in \cref{eq:lindblad-expansion}. As derived in \cref{app:1storder}, the first-order terms are given by
\begin{align}\label{eq:first-order-solution}
\mathcal P_{\mathbf m}^{(1)}(t,t_0) = \sum_k \gamma_k \!\sum_{\mathbf n \leq \mathbf m} \int_{t_0}^t \!\!\mathcal P_{\mathbf n}^{(0)}(t,t') \mathcal L_k^\mathrm{err} \mathcal P_{\mathbf m-\mathbf n}^{(0)}(t',t_0)\, \mathrm dt'\!.
\end{align}
This expression holds independently of the simplifying assumptions made in the previous section and solely relies on having access to a computable representation of $\mathcal P^{(0)}_{\mathbf m}(t,t_0)$ (which can be obtained by solving \cref{eq:ZPG} recursively using $\mathcal{L}^\mathrm{ideal}_\mathrm{ZPG}$). The map $P_{\mathbf m}^{(1)}(t,t_0)$ describes the first-order correction conditioned on the measured photon pattern $\mathbf m$ and can be intuitively understood as the summation over all possibilities of having undergone ideal evolution while producing the photon pattern $\mathbf m-\mathbf n$ up until some time $t'$, experiencing an error ($\mathcal L_k^\mathrm{err}$), and then evolving ideally again while producing the remaining photon pattern $\mathbf n$.

Since the zero-order conditional maps entering \cref{eq:first-order-solution} have the closed form in \cref{eq:zero-order-closed-form}, the first-order error corrections can be computed directly. We report higher-order perturbations in terms of zero-order solutions in \cref{sec:higher-order}. However, we highlight here that the computation of the $k$-th order solution involves solving a $k$-th order integral.

With these perturbative expansions, we can readily compute the Choi matrices for any conditional map $\mathcal P_{\mathbf m}$.
In the next section, we show how overlaps with these matrices yield gate fidelity and the Pauli error weights to the desired perturbative order, conditioned on photon detection patterns.

\subsection{Fidelity and Pauli error weights}\label{sec:fidelity}
We now apply our results to the derivation of quantities of interest in quantum computing. This solidifies the connection between quantum dynamics, which captures the effects of hardware errors, and quantum information. In particular, we calculate expressions for fidelity and Pauli error weights. As described in \cref{app:general-fidelity}, our method applies to general Lindbladian error terms. However, many relevant sources of error can be described using terms of the following form
\begin{align}\label{eq:incoherent-errors}
\mathcal L^{\mathrm{err}} = \sum_{j} \gamma_j \mathcal D(L_j).
\end{align}
This represents an incoherent error with a collapse operators $L_j$. As we shall demonstrate, errors of this form produce insightful representations of our quantities of interest, further motivating the restriction to this case.

\paragraph*{Fidelity} In many applications, such as quantum networks, one is often interested in preparation of a particular state (e.g., a Bell pair). In such cases, it is important to know how close the final state of the protocol is to the desired state for a particular input state, say $\ket{\psi_{\mathrm{in}}}$. We denote the target state that we achieve by post-selecting on PNRD measurement $\mathbf m$, by $\ket{\psi_{\mathbf m}}$. Hence, the quantity of interest is
\begin{align}
F^{\mathrm{state}}_{\mathbf m} = \frac{1}{p_{\mathbf m}}\bra{\psi_{\mathbf m}} \mathcal P_{\mathbf m}[\ket{\psi_{\mathrm{in}}}\bra{\psi_{\mathrm{in}}}] \ket{\psi_{\mathbf m}},
\end{align}
with $p_{\mathbf m} = \mathrm{tr}(\mathcal P_{\mathbf m}[\ket{\psi_{\mathrm{in}}}\bra{\psi_{\mathrm{in}}}])$ being the success probability. We also use $p_{\mathbf m}^{(0)}$ to denote the success probability in the ideal (no noise) case. Building on our perturbative expansions, in \cref{app:fidelity-expansion} we obtain
\begin{align}\label{eq:fid-main}
\begin{split}
F_{\mathbf m}^{\mathrm{state}} &= 1-\frac{1}{p_{\mathbf m}^{(0)}} \sum_{j} \gamma_j \\
&\quad \sum_{\mathbf n\le \mathbf m} \int_{t'=0}^t \left\|\Pi_{\mathbf m}^\perp K_{\mathbf n}(t,t') L_j K_{\mathbf m-\mathbf n}(t',0) \ket{\psi_{\mathrm{init}}}\right\|^2\, \mathrm dt'\\
&+ O\left( (\sum_{j}\gamma_j)^2 \right),
\end{split}
\end{align}
where $K_{\mathbf n}(t_2,t_1) := \sqrt{f_{\mathbf n}(t_2-t_1)} e^{-iH_{\mathrm{eff}}(t_2-t_1)} \mathbf D^{\mathbf n}$, and where $\Pi_{\mathbf m}^\perp : = \mathbb I - \ket{\psi_{\mathbf m}}\bra{\psi_{\mathbf m}}$ is the projection operator on to the subspace orthogonal to the target state.

\paragraph*{Pauli error weights} 
When designing photon-heralded gates, one is often interested in calculating Pauli error weights, and measures such as the `average gate fidelity,' or the `minimum fidelity' \cite{nielsen2010quantum}. Denoting the set of successful outcomes of PNRD measurement by $A \subset \mathbb N^N$, and the inverse of the target gate of heralding on $\mathbf m$ by $R_{\mathbf m}$, we establish, in \cref{sec:pauli}, that the probability of Pauli error $P$ occurring during the noisy gate is given by

\begin{align}\label{eq:pauli-err-main}
\begin{split}
\chi_{P,P} &= \frac{1}{2^{2n} p_A^{(0)}} \sum_{\mathbf m\in A} \sum_{\mathbf n\le \mathbf m} \sum_j \gamma_j\\
&\quad\int_{t'=0}^t \left|  \mathrm{Tr}\left( P\, R_{\mathbf m} K_{\mathbf n}(t,t') L_j K_{\mathbf m-\mathbf n}(t',0) \right) \right|^2\, \mathrm dt'\\
&+O\left( (\sum_j \gamma_j)^2 \right),
\end{split}
\end{align}
for the Pauli errors $P\ne \mathbb I$. Lastly, the entanglement fidelity can be calculated as $F_{\mathrm{ent}} = 1-\sum_{P\neq \mathbb I} \chi_{P,P}$, and that the average gate fidelity is related to the entanglement fidelity via $F_{\mathrm{avg}} = (2^n F_{\mathrm{ent}} + 1)/(2^n +1)$, and hence, these quantities can be calculated efficiently by our approach.

Mappings such as \cref{eq:fid-main} and \cref{eq:pauli-err-main}, from the physical hardware error mechanisms to quantum information measures of imperfection, are a key result of our work. As mentioned above, with such mappings, one can perform more reliable threshold simulations. 

Our fidelity and Pauli error formulas (\cref{eq:fid-main} and \cref{eq:pauli-err-main}) have simple physical interpretations: they are imperfections (either the overlap with error subspace $\Pi_{\mathbf m}^\perp$, or the overlap with the Pauli operator) that our system accumulates in evolving under a partial conditional map described by $K_{\mathbf m - \mathbf n}$, incurring an error $L_j$, and then evolving with some other conditional map (described by $K_{\mathbf n}$) that results in having $\mathbf m$ photons overall. We also observe an interesting phenomenon that the contributions corresponding to anticommutators $\{L^\dag_j L_j,\bullet\}$ in $\mathcal D(L_j)$ do not appear in expressions \cref{eq:fid-main} and \cref{eq:pauli-err-main}. In other words, to first order, these anti-commutator terms only contribute to altering the success probability of the gate, and not the quality of the gate conditioned on a success. We also refer to \cref{eq:pert-in-prob} for the perturbative formulation of the success probabilities.

\paragraph*{Computational advantages} Our approach can be used to compute errors numerically, not just analytically. This is especially useful if the integrals in \cref{eq:first-order-solution} are hard to evaluate analytically. We highlight that such a numerical calculation requires $O(d^3)$ time, whereas the direct numerical simulation of the ZPG evolution in the density matrix representation requires $O(d^6)$. This is because our zero-order solutions produce unnormalized state vectors, and the calculation of quantities such as fidelity and Pauli error reduces to computation of the overlaps of the form $\bra{\psi} A_1\cdots A_k \ket{\psi}$ or $\mathrm{Tr}(A_1\cdots A_k)$ for $k=O(1)$. Operators $A_1,\cdots, A_k$ are each computable in time $O(d^3)$. Given that matrix-vector multiplication and matrix-matrix multiplication have complexity $O(d^2)$ and $O(d^3)$ respectively, we require an overall time complexity $O(d^3)$. To further elaborate, the operators $A_i$ consist of exponential of the effective Hamiltonian ($\exp(iH_{\mathrm{eff}}t)$ and multiplications of $\mathbf D$ gates which takes $O(d^3)$ to compute in worst case).

The phenomenon of runtime reduction in perturbative frameworks was also noted in \cite{karimi2026comparing}.

\section{Application of the method}\label{sec:application}
Our method applies broadly to probabilistic entanglement-generation schemes, including the single-photon heralding protocol \cite{cabrillo1999creation} and the two-photon Barrett--Kok protocol \cite{barrett2005efficient}. As a simple example, in \cref{sec:n-protocol} we calculate the perturbative conditional states for the heralded single-photon protocol and show agreement with the exact analytical results of Ref.~\cite{wein2020analyzing}.
The framework also extends to measurement-based, photon-mediated two-qubit gates, including the repeat-until-success (RUS) $\mathsf{CZ}$ gate \cite{lim2005repeat,lim2006repeat,de2024spin} and the $\mathsf{CNOT}$ teleportation gate \cite{simmons2024scalable}, where it can be used to characterize conditional gate operations.

Motivated by the need to analyze gate designs for distributed quantum computing, we use our method to derive perturbative conditional maps for the RUS $\mathsf{CZ}$ gate \cite{lim2005repeat}. For quantum error correction, the important quantities extracted from these conditional maps are the induced Pauli error weights, computed from the corresponding process matrix via \cref{eq:choi} or \cref{eq:pauli-err-main}. In the following, we apply our framework to derive these error weights for several sources of imperfection in the RUS gate. We first study spin-related noise sources, and then consider coherent errors in the implementation of the linear-optical transformation. We compare these analytical results with exact numerical simulations using the Python package ZPGenerator \cite{quandela_zpgenerator_2024}.

While some Pauli error weights for this gate have been previously studied \cite{de2024spin, chan2026practical}, our perturbative method connects those qubit level errors directly to hardware-level imperfections and enables systematic extensions to more detailed physical noise models.

\begin{figure}[t]
\centering

\begin{minipage}{\linewidth}
    \centering
    \includegraphics[width=\linewidth]{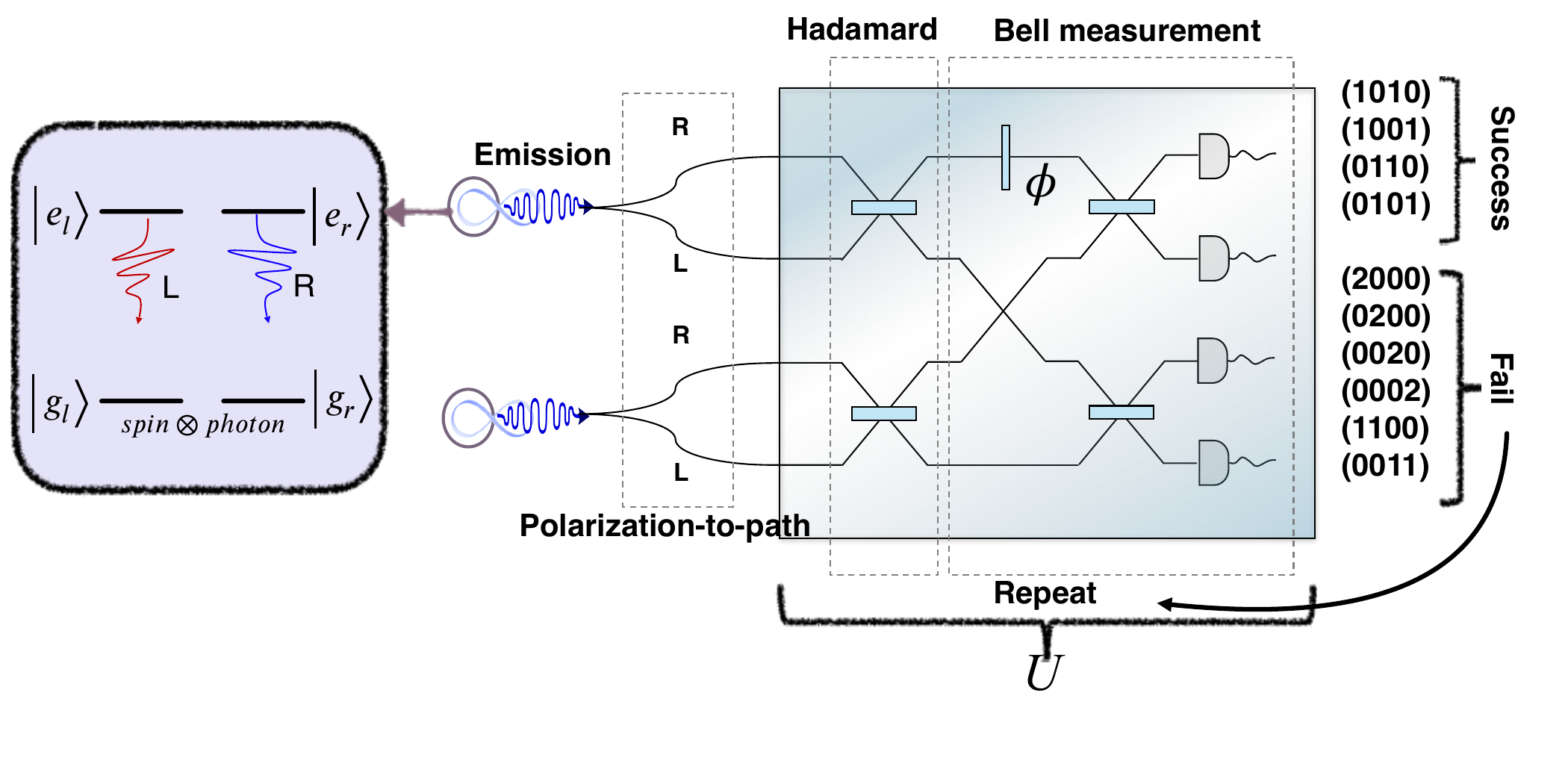}

    \vspace{0.01em}
    \panellabel{fig:RUS-levels-and-table-a}
\end{minipage}

\vspace{0.8em}

\begin{minipage}{0.8\linewidth}
    \centering
    \footnotesize
    \setlength{\tabcolsep}{4pt}
    \renewcommand{\arraystretch}{1.15}

    \resizebox{\linewidth}{!}{%
    \begin{tabular}{|c|c|c|}
    \hline
    \textbf{Pattern ($m$)} & \textbf{$D_iD_j$} & \textbf{Resulting Gate} \\
    \hline
    1010 & $D_1 D_3$ & $-e^{i\pi/4}(S\!\otimes\! S^\dagger)\,\mathsf{CZ}$ \\
    \hline
    1001 & $D_1 D_4$ & $-e^{-i\pi/4}(S^\dagger\!\otimes\! S)\,\mathsf{CZ}$ \\
    \hline
    0110 & $D_2 D_3$ & $e^{-i\pi/4}(S^\dagger\!\otimes\! S)\,\mathsf{CZ}$ \\
    \hline
    0101 & $D_2 D_4$ & $e^{i\pi/4}(S\!\otimes\! S^\dagger)\,\mathsf{CZ}$ \\
    \hline
    \end{tabular}%
    }

    \vspace{0.2em}
    \panellabel{fig:RUS-levels-and-table-b}
\end{minipage}
\caption{(a)
A degenerate level structure of an emitter with spin-selective polarization emission for implementing the two-qubit RUS gate, together with the corresponding photonic integrate circuit implementing a dual-rail Bell-state measurement in the photonic X basis. Here, $U$ denotes the interferometer unitary, defined in \cref{eq:U-of-rus}. (b) A list the successful photon-detection patterns and the corresponding spin qubit operations produced by the RUS gate. Here $S=\mathrm{diag}(1,i)$ (and $S^
{\dagger}$) is a required local Clifford correction. The operators $D_i$ are linear combinations of lowering operators associated with different photon emitter transitions, where the coefficients are determined by the interferometer network (cf. \cref{eq:def-of-D}). 
}
\label{fig:RUS-levels-and-table}
\end{figure}

\paragraph{Protocol description} 
The RUS gate is introduced in \cite{lim2005repeat} to realize  a quasi-deterministic $\mathsf{CZ}$ gate between non-interacting qubits. We restate the protocol using the terminology introduced in \cref{sec:method}, focusing on an idealized four-level emitter structure representative of a charged quantum dot. An implementation with two quantum emitters and the corresponding linear-optical entangling circuit is illustrated in \cref{fig:RUS-levels-and-table}\ref{fig:RUS-levels-and-table-a}. Each emitter has two ground states and two excited states, with the optical transitions emitting into orthogonal polarization modes. Although we use an idealized model to demonstrate the method, the same framework can incorporate more realistic effects such as interactions with the nuclear bath and spin precession in an external magnetic field.

To implement the gate, each source emits a photon, which is then routed into an interferometer implementing the unitary $U$ defined by
\begin{align}\label{eq:U-of-rus}
\resizebox{\columnwidth}{!}{$
U = \begin{pmatrix}
\cos^2\theta & \cos\theta\sin\theta & \cos\theta\sin\theta & \sin^2\theta\\
\cos\theta\sin\theta & \sin^2\theta & -\cos^2\theta & -\cos\theta\sin\theta\\
\cos\theta\sin\theta & -\cos^2\theta & e^{i\phi}\sin^2\theta & -e^{i\phi}\cos\theta\sin\theta\\
\sin^2\theta & -\cos\theta\sin\theta & -e^{i\phi}\cos\theta\sin\theta & e^{i\phi}\cos^2\theta
\end{pmatrix}
$}
\end{align}
which implements a probabilistic Bell-state measurement in the X basis of the emitted photons.

The measurement outcomes either herald success or failure. Success outcomes implement a two-qubit unitary on the ground state qubits of the emitters equivalent to 
$\mathsf{CZ}$ up to local single-qubit operations. In the ideal setting—neglecting photon loss, dephasing, and other imperfections—the failure outcomes implement a unitary that is, again up to local operations, equivalent to the identity. This feature allows the operation to be repeated without reinitializing the qubits until a success outcome is obtained. The set of success-conditioned unitary gates acting on the spin qubits is summarized in \cref{fig:RUS-levels-and-table}\ref{fig:RUS-levels-and-table-b}.

\paragraph{Zero-order analysis} 
In the rotating frame of the emitters, the effective Hamiltonian, corresponding to $\mathcal L_{\mathrm{ZPG}}^\mathrm{ideal}$ is given by
\begin{align}
\begin{split}\label{eq:H_RUS}
H_{\mathrm{eff}} = -i\frac\gamma2 ( &\ket{e_{\ell,1}}\bra{e_{\ell,1}}+ \ket{e_{r,1}}\bra{e_{r,1}}\\
&+\ket{e_{\ell,2}}\bra{e_{\ell,2}} + \ket{e_{r,2}}\bra{e_{r,2}}),
\end{split}
\end{align}
where $\gamma$ is the decay rate of an emitter.

For the ideal case, we set $\theta = \frac\pi4$, $\phi = \frac\pi2$.
The detector operators are given by $\mathbf D = U \boldsymbol \sigma$, which, in the ideal case yields
\begin{align}
\begin{split}
D_1 &= \frac12 \left( \sigma_{r,1} + \sigma_{\ell,1} + \sigma_{r,2} + \sigma_{\ell,2} \right)\\
D_2 &= \frac12 \left( \sigma_{r,1} + \sigma_{\ell,1} - \sigma_{r,2} - \sigma_{\ell,2} \right)\\
D_3 &= \frac12 \left( \sigma_{r,1} - \sigma_{\ell,1} +i \sigma_{r,2} -i \sigma_{\ell,2} \right)\\
D_4 &= \frac12 \left( \sigma_{r,1} - \sigma_{\ell,1} -i \sigma_{r,2} +i \sigma_{\ell,2} \right)
\end{split}
\end{align}
where $\sigma_{k,a} = \ket{g_{k,a}}\bra{e_{k,a}}$ for $k\in\{\ell,r\}$ and $a\in\{1,2\}$ labeling polarization and emitter respectively. Notably, the zero-order conditional maps $
\mathcal P_{\mathbf m}$ for $|\mathbf m|=2$ are proportional to unitary maps on the computational ground-state subspace. Depending on the detection pattern, these maps are either proportional to a tensor product of local gates on individual emitters or, up to local operations, proportional to the $\mathsf{CZ}$ gate. 
For these two-photon branches, the jump product maps the initially excited population into the ground-state computational subspace, so the remaining no-jump evolution acts trivially on the resulting spin-qubit operation: $\exp(iH_{\mathrm{eff}}t) D_i D_j = D_{i} D_j$. As such, the ideal action of the gate is influenced solely by the detector operators.

\paragraph{Comparison with numerical simulation}  The zero-order analysis reproduces the action of the RUS gate within our formalism. We now perturbatively include noise and analyze its impact on the conditional gate operation. Following the techniques described in \cref{sec:fidelity}, we compute the diagonal process-matrix elements $\chi_{P,P}$, the Pauli error weights, for a noisy RUS gate conditioned on the success pattern $\mathbf m = (1010)$ and also the failure pattern $\mathbf m = (2000)$. 

As representative emitter-related imperfections, we consider spin relaxation in the excited state at rate $\gamma_1$, spin dephasing in the excited state at rate $\gamma_2$, and emitter optical dephasing between ground and excited states at rate $\gamma_3$. For each case, we compute the resulting Pauli error weights of the RUS gate. In the analytical results shown below, we consider perturbative terms up to at most $4$-th order, using the extension described in \cref{sec:higher-order}.

\begin{figure*}[t!]
\centering

\begin{minipage}[t]{0.335\textwidth}
    \centering
    \includegraphics[width=\linewidth]{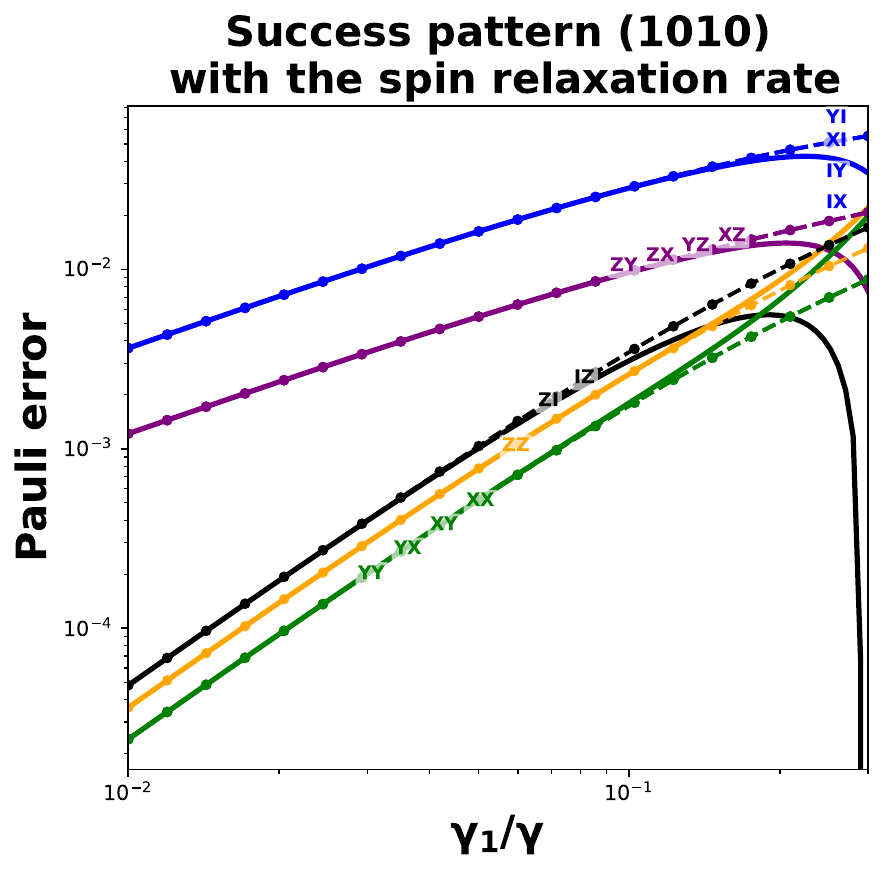}
    
    \vspace{0.2em}
    \panellabel{fig:rus-a}
\end{minipage}\hfill
\begin{minipage}[t]{0.31\textwidth}
    \centering
    \includegraphics[width=\linewidth]{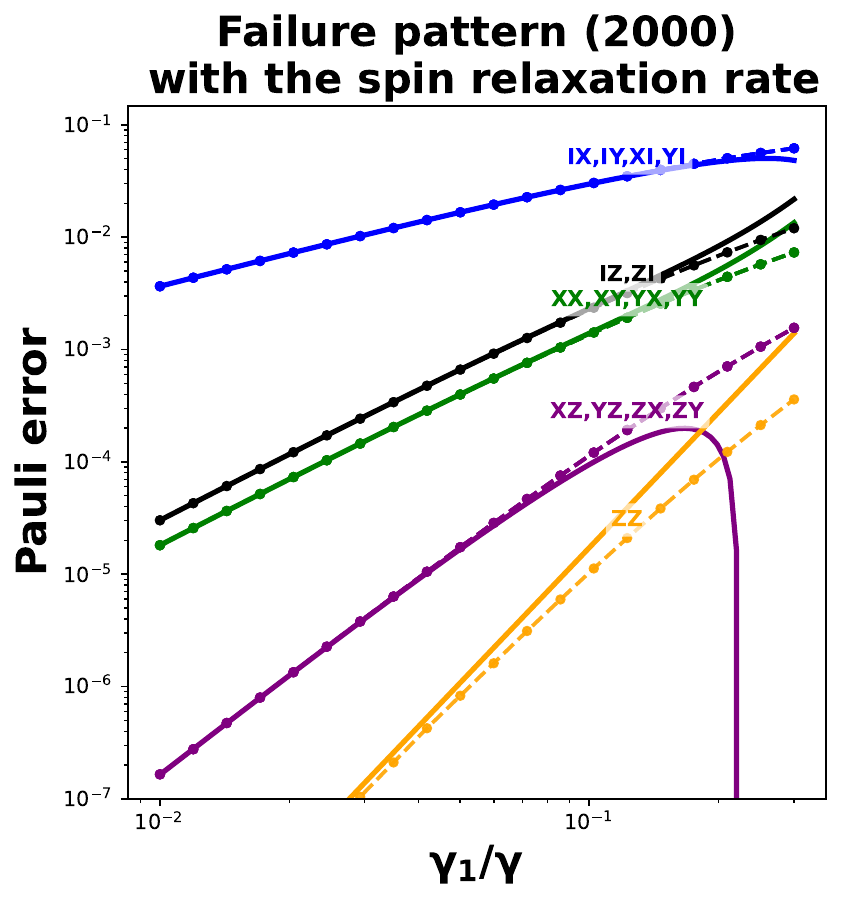}
    
    \vspace{0.2em}
    \panellabel{fig:rus-b}
\end{minipage}\hfill
\begin{minipage}[t]{0.335\textwidth}
    \centering
    \includegraphics[width=\linewidth]{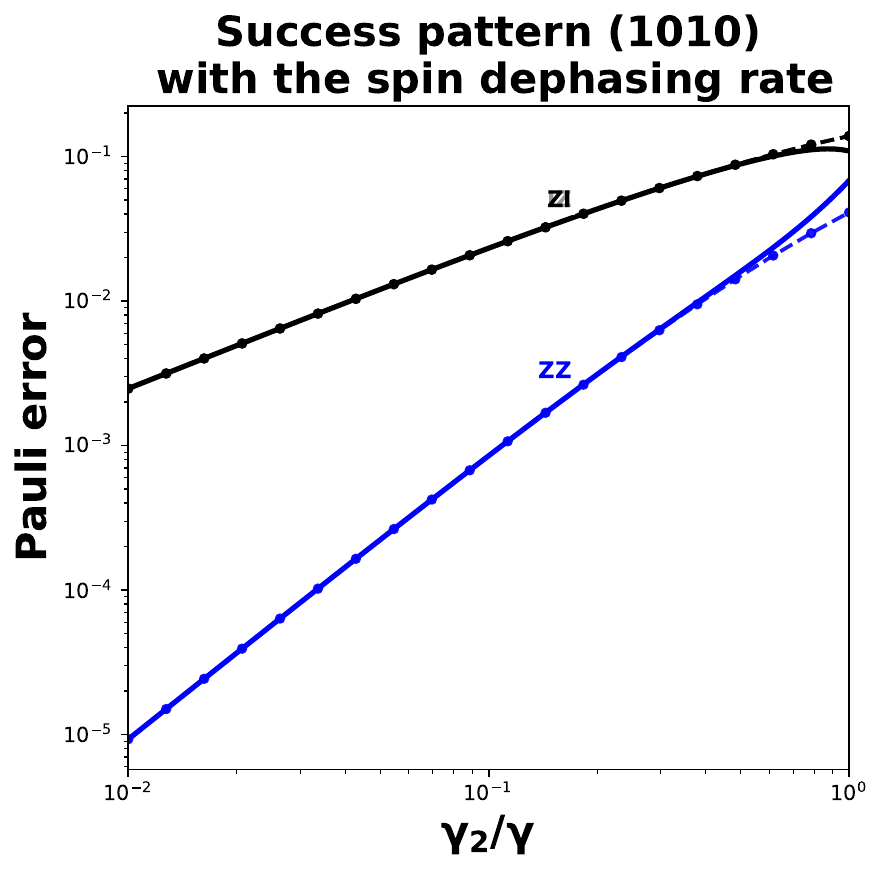}
    
    \vspace{0.2em}
    \panellabel{fig:rus-c}
\end{minipage}

\vspace{1em}

\hspace*{\fill}
\begin{minipage}[t]{0.31\textwidth}
    \centering
    \includegraphics[width=\linewidth]{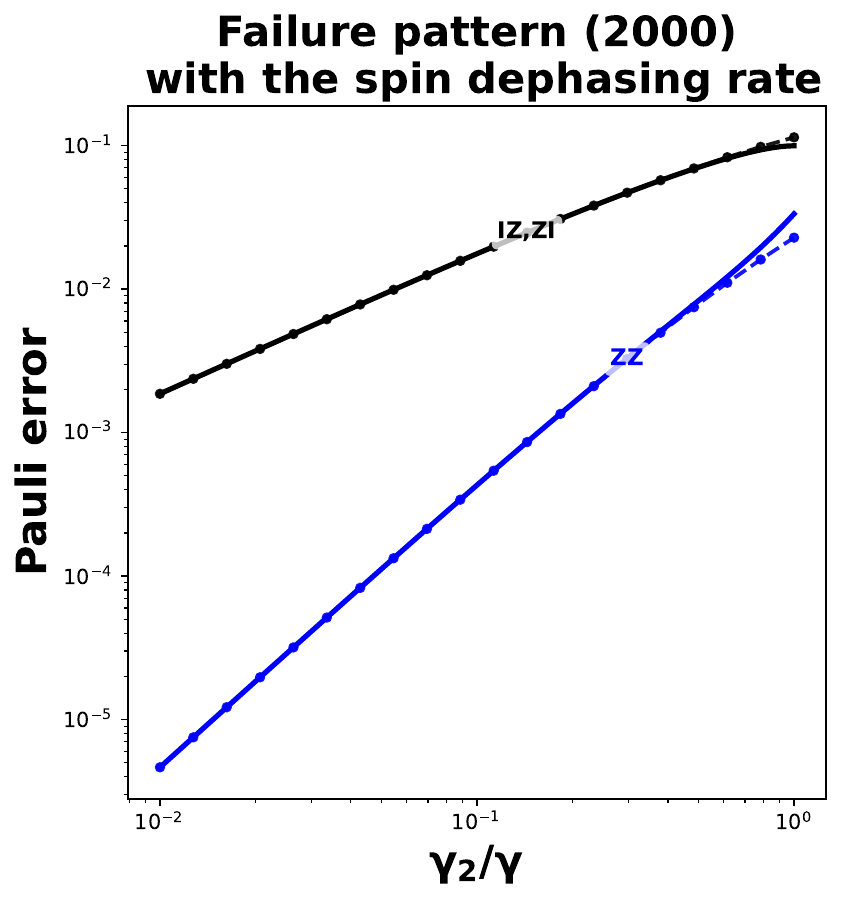}
    
    \vspace{0.2em}
    \panellabel{fig:rus-d}
\end{minipage}
\hspace{0.06\textwidth}
\begin{minipage}[t]{0.335\textwidth}
    \centering
    \includegraphics[width=\linewidth]{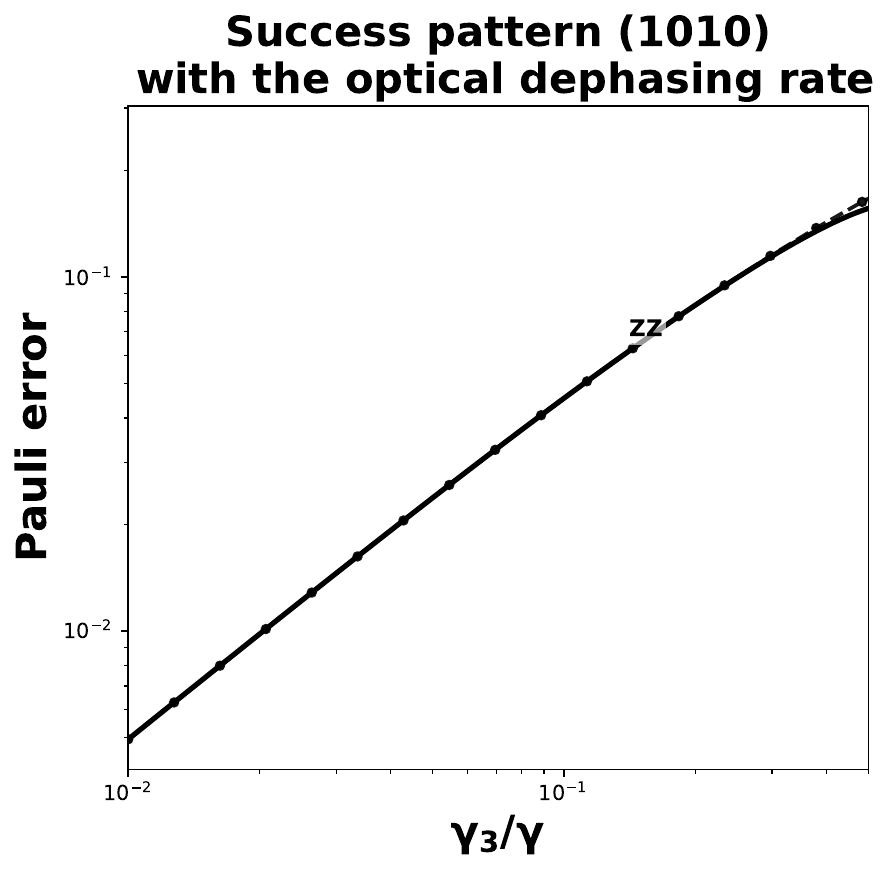}
    
    \vspace{0.2em}
    \panellabel{fig:rus-e}
\end{minipage}
\hspace*{\fill}

\vspace{1.2em}

\begin{minipage}[t]{0.98\textwidth}
\centering
\fontsize{6.5}{8.4}\selectfont
\setlength{\tabcolsep}{4pt}

\resizebox{\textwidth}{!}{%
\begin{tabular}{ll}
\hline\hline
\multicolumn{2}{c}{\textbf{Successful pattern $\mathbf m=(1,0,1,0)$}} \\
\hline
\textbf{Pauli error} & \textbf{Error probability} \\
\hline
$IX, IY, XI, YI$ & $\frac38(\gamma_1/\gamma) - \frac98(\gamma_1/\gamma)^2 - \frac{3}{32} (\gamma_1\gamma_2/\gamma^2)$ \\
$IZ, ZI$ & $\frac14(\gamma_2/\gamma) + \frac12(\gamma_1/\gamma)^2 -\frac34(\gamma_1\gamma_2/\gamma^2) -\frac3{16} (\gamma_2/\gamma)^2$ \\
$XX, XY, YX, YY$ & $\frac14 (\gamma_1/\gamma)^2$ \\
$XZ, YZ, ZX, ZY$ & $\frac18 (\gamma_1/\gamma) + \frac1{8}(\gamma_3\gamma_1/\gamma^2)-\frac38(\gamma_1/\gamma)^2 + \frac3{32} (\gamma_1\gamma_2/\gamma^2)$ \\
$ZZ$ & $\frac12(\gamma_3/\gamma) - \frac12 (\gamma_3/\gamma)^2 -\frac32(\gamma_3\gamma_1/\gamma^2) - \frac38 (\gamma_3\gamma_2/\gamma^2) + \frac38(\gamma_1/\gamma)^2 + \frac3{32} (\gamma_2/\gamma)^2$ \\
\hline
\multicolumn{2}{c}{\textbf{Failure pattern $\mathbf m=(2,0,0,0)$}} \\
\hline
$IX, IY, XI, YI$ & $\frac38(\gamma_1/\gamma) - \frac{15}{16}(\gamma_1/\gamma)^2$ \\
$IZ, ZI$ & $\frac{3}{16}(\gamma_2/\gamma) + \frac{5}{16}(\gamma_1/\gamma)^2- \frac{15}{32}(\gamma_1\gamma_2/\gamma^2) - \frac{15}{128}(\gamma_2/\gamma)^2$ \\
$XX, YY, XY, YX$ & $\frac3{16}(\gamma_1/\gamma)^2$ \\
$XZ, ZX, YZ, ZY$ & $\frac3{32}(\gamma_1\gamma_2/\gamma^2)$ \\
$ZZ$ & $\frac{3}{64}(\gamma_2/\gamma)^2$ \\
\hline\hline
\end{tabular}%
}

\vspace{0.3em}
\panellabel{fig:rus-f}
\end{minipage}

\caption{
Demonstration of the Pauli error weights associated with different sources of noise. Here $\gamma_1$ denotes the excited-state spin relaxation noise strength defined in ~\cref{eq:thremalization-def}, $\gamma_2$ and $\gamma_3$ are the spin dephasing rate and emitter optical dephasing rate, respectively (see \cref{eq:spin-dephasing-def,eq:optical-dephasing-def}), and $\gamma$ is the optical decay rate defined in \cref{eq:H_RUS}. Panels (\ref{fig:rus-a})--(\ref{fig:rus-e}) showcase the effect of each individual error, when the rest are set to zero. In panels (\ref{fig:rus-a})--(\ref{fig:rus-e}), the analytical expression (represented by the solid line) corresponds to the solution that we obtain with our approach up to the \textit{fourth order}. The dashed-dotted lines represent the numerical solutions.
We highlight that missing curves indicate that the missed Pauli error is identically zero. Furthermore, the failure pattern for the optical dephasing error ($\gamma_3)$, had no errors, and hence, we have only plotted errors associated with the successful pattern only (panel (\ref{fig:rus-e})).
Panel (\ref{fig:rus-f}) lists the corresponding analytical Pauli error probabilities for the successful pattern $\mathbf m=(1,0,1,0)$ and the failure pattern $\mathbf m=(2,0,0,0)$.
Errors are computed to second order in $(\sum_i\gamma_i)/\gamma$ in the limit $t\to\infty$. We have chosen to present the table to the second-order for the simplicity in presentation, while the arbitrary-order solution can also be obtained using the provided Mathematica notebook \cite{karimi_perturbative_zpg_code}.}
\label{fig:rus-full-two-column}
\end{figure*}

For completeness, we also present the combined-error scenario—where all three noise sources are present—in \cref{fig:rus-full-two-column}\ref{fig:rus-f}.
Finally, we apply the expression derived in \cref{sec:fidelity} to a coherent photonic-circuit error, namely an interferometer phase-shifter miscalibration.

For the incoherent errors considered, we restrict the error Lindbladians to the excited subspace. This allows us to take the limit $t\to\infty$, corresponding to waiting until the optical emission process is complete. If ground-state errors were included, the conditional error channel would continue to evolve after photon emission, and the result would depend on the chosen waiting time. Such effects can still be treated within our framework by keeping a finite final time $t$, but we omit them here to avoid introducing additional hardware-dependent timing assumptions. 

\textbf{1. Excited state spin relaxation.} We first consider excited-state spin flips occurring at rate $\gamma_{1}$. The corresponding error Lindbladian is
\begin{align}\label{eq:thremalization-def}
\mathcal L_{1}^\mathrm{err} = \mathcal D(\ket{e_\ell}\bra{e_r}) + \mathcal D(\ket{e_r}\bra{e_{\ell}}),
\end{align}
acting independently on each quantum emitter. Substituting this error generator into the perturbative expansion \cref{eq:first-order-solution}, and then using the Choi-matrix construction in \cref{eq:choi}, gives the corresponding diagonal Pauli error weights.
We compare these analytical estimates with the exact numerical simulations in \cref{fig:rus-full-two-column}\ref{fig:rus-a}, \ref{fig:rus-b}, finding excellent agreement for $\gamma_1/\gamma < 0.1$, which is the relevant regime for most quantum error correction codes.

\textbf{2. Excited state spin dephasing.} We next consider dephasing within the excited-state manifold. The corresponding error generator is
\begin{align}\label{eq:spin-dephasing-def}
\mathcal L_{2}^\mathrm{err} = \mathcal D(\sigma_z),
\end{align}
where
$\sigma_z:=\frac12\left(\ket{e_\ell}\bra{e_\ell} - \ket{e_r}\bra{e_r}\right)$, acting independently on each emitter.
Following the same procedure as for spin relaxation, we compute the Pauli error weights and compare them with numerical simulations in \cref{fig:rus-full-two-column}\ref{fig:rus-c}, \ref{fig:rus-d}, again finding excellent agreement for $\gamma_2<0.4$, which is more than sufficient to analyze most quantum error correction codes.

\textbf{3. Pure optical dephasing.} The error generator corresponding to the emitter optical dephasing rate of $\gamma_3$ is
    \begin{align}\label{eq:optical-dephasing-def}
    \mathcal L_{3}^\mathrm{err} = \mathcal D(\ket{e_r}\bra{e_r} + \ket{e_\ell}\bra{e_\ell}).
    \end{align}
    Using the perturbative expansion, we find that the Pauli error weights contain only one non-zero term, apart from the identity contribution, provided as
\begin{align}
\begin{split}
(\chi_{(1010)})_{ZZ} &= \frac12 \frac{\gamma_3}\gamma - \frac12 \left(\frac{\gamma_3}\gamma\right)^2 + O\left(\left(\frac{\gamma_3}\gamma\right)^3\right)
\end{split}
\end{align}
in the limit of $t\to\infty$. This error model prohibits errors other than Z \cite{de2024spin}.

As before, we compare the perturbative prediction with numerical simulations in \cref{fig:rus-full-two-column}\ref{fig:rus-e}. We find that emitter optical dephasing does not affect the conditional channel associated to $\mathbf m = (2000)$ in either the analytic estimate or the exact numerical solution. This is consistent with the previous analysis in \cite{de2024spin}. It also illustrates a useful feature of polarization-encoded spin-photon entangled emission over alternative schemes, such as time-bin encoding \cite{chan2026practical}.

\setcounter{panel}{0}

We extend this analysis further by considering the combined effect of all the above noise sources. The result of this analysis is provided in \cref{fig:rus-full-two-column}\ref{fig:rus-f}. This captures the interplay between noise parameters through higher-order cross terms (e.g., $\gamma_i\gamma_j$ contributions in \cref{fig:rus-full-two-column}\ref{fig:rus-f}). Such analytic expressions are useful for modeling combined errors and identifying multi-parameter trends without relying on exhaustive numerical sweeps, whose cost grows rapidly with the number of independent noise parameters.

\begin{figure}[t] \centering \begin{minipage}[b]{\linewidth} \centering \includegraphics[width=\linewidth]{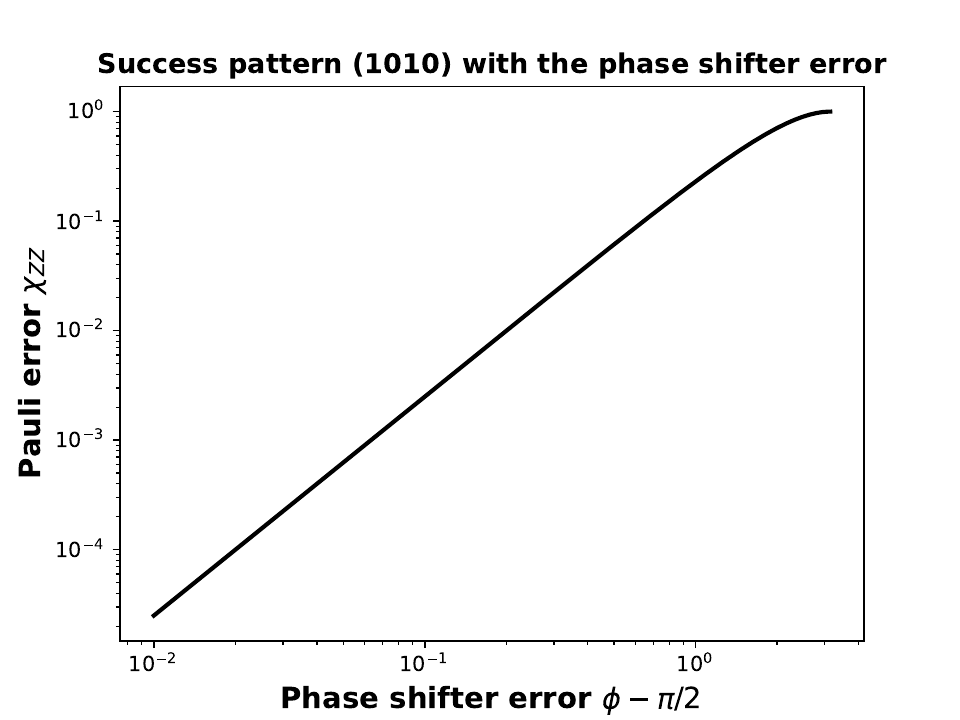} \vspace{0.5ex} \textbf{(a)} \end{minipage} \hfill \begin{minipage}[b]{\linewidth} \centering \renewcommand{\arraystretch}{1.2} \begin{tabular}{ll} \hline\hline \multicolumn{2}{c}{\textbf{Successful pattern $\mathbf m=(1,0,1,0)$}} \\ \hline \textbf{Pauli error} & \textbf{Error probability} \\ \hline $ZZ$ & $\sin^2(\delta/2)$ \\ all the rest & 0\\ \hline \multicolumn{2}{c}{\textbf{Failure pattern $\mathbf m=(2,0,0,0)$}} \\ \hline all errors & $0$\\ \hline\hline \end{tabular} \vspace{0.5ex} \textbf{(b)} \end{minipage} \caption{The Pauli errors associated with the phase shifter error in the RUS gate. (a) The successful pattern $(1010)$ has only a nonzero diagonal $ZZ$ Pauli error weight, while the failure pattern $(2000)$ remains unaffected by this imperfection. (b) Corresponding error table for the successful pattern.} \label{fig:phase-shifter} \end{figure}

\textbf{4. Phase shifter error.}
The framework also captures coherent errors and errors that affect the emitted light only. As an example of this, we study the effects of phase shifter miscalibration error (i.e., $\phi = \frac\pi2 + \delta$ in \cref{eq:U-of-rus}). Our treatment for this error is different from the previous errors, since it comes as a perturbation to $\mathbf D$ operators rather than the Lindbladian. Following the calculation detailed in \cref{sec:phase-shifter}, we obtain that the only non-zero diagonal Pauli error weights associated with the conditional map $(1010)$ is
\begin{align}
(\chi_{(1010)})_{ZZ} = \sin^2\left( \frac{\delta}{2} \right).
\end{align}
We also obtain off-diagonal $II,ZZ$ and $ZZ,II$ entries as $\chi_{II,ZZ} = \chi_{ZZ,II} = -\frac12\sin\delta$, these elements are indicators of the coherent nature of the underlying phase shift error. The ability of our method to capture these elements show how it is capable of providing insight beyond the standard assumption of stochastic Pauli noise. 
As discussed in \cref{sec:phase-shifter}, and similar to optical dephasing, the conditional map $(2000)$ incurs no error due to the phase shifter imperfection. We refer to \cref{fig:phase-shifter} for the plot.

\section{summary and outlook}\label{outlook}

We developed a perturbative framework for analyzing errors in quantum operations based on heralded light--matter interactions. As a representative application, we derived Pauli error weights for hardware-level error mechanisms impacting a repeat-until-success CZ gate. This provides a direct connection between microscopic imperfections and effective qubit-level noise models for hybrid light-matter platforms. While the theoretical and numerical results agree well in the low-noise regime, numerical simulations become increasingly delicate when extracting very small error probabilities, where finite precision may introduce artifacts. By contrast, our analytical expressions remain valid in this regime and provide reliable predictions, making them a useful complement to numerics. In addition, these expressions provide a clear breakdown of the dominant error types, which is useful for designing hardware-aware error-correction strategies.

An advantageous aspect of our work is the ability to capture the interplay between different error mechanisms. For instance, \cref{fig:rus-full-two-column}\ref{fig:rus-f} contains cross term of the form $
\gamma_i\gamma_j$, which quantify how distinct noise sources combine beyond their independent contributions. Such analytic mappings can be used in fault-tolerance threshold studies, enabling systematic exploration of parameter regimes, efficient tuning of device specifications, and clear quantification of trade-offs among competing noise sources. This is particularly useful because exploring large parameter spaces with numerical tools alone becomes costly due to the curse of dimensionality.

One limitation of the ZPG framework, in both simulation and theory, is that it does not directly support time-bin single-qubit bit flip gates because the present formulation is effectively local in time. It can capture instantaneous unitary transformations to time modes emitted synchronously by one or more sources, but it cannot yet represent full unitary mixing between time-delayed modes.  In its present form, this method can model time-bin qubit generation and phase-type operations as in \cite{wein2020analyzing}, but not full Bloch-sphere control of those emitted time bin qubits. Capturing such operations would require the ZPG framework to include explicit nonlocal-in-time mode transformations, or temporal mode mixing, within an effective source evolution.

Furthermore, obtaining closed-form expressions requires the analytic evaluation of low-order integrals (e.g., \cref{eq:first-order-solution}). 
These integrals may not always admit a convenient analytic form. In such cases, the same perturbative coefficients can instead be computed numerically, while still avoiding numerically simulating the full ZPG dynamics. This hybrid analytic-numerical approach can provide useful insight into the structure of the errors. Our approach also facilitates composition of conditional error channels. For example, one can compute an average error channel for a RUS gate after a number of repetitions.
While the method presented here can incorporate arbitrary Markovian imperfections, we leave non-Markovian extensions to future work.

Although we focused on unitary gates as an example, the methodology is broadly applicable to conditional process maps arising from open-system dynamics and photon-detection protocols, including measurements and other non-unitary operations.
Examples include Bell-state measurements in entanglement-generation protocols involving spin-photon entanglement and heralded remote spin-spin entanglement, parity-check operations in quantum error correction, and weak measurements.
This technique applies to experimental platforms that realize entanglement between non-interacting emitters, such as trapped ions \cite{moehring2007entanglement} and neutral atoms \cite{hofmann2012heralded}, and is especially valuable for solid-state systems, where emitter-level noise is often significant.
We anticipate that this framework will support threshold studies of realistic physical hardware and provide design guidance for scalable quantum technologies, ranging from large-scale quantum computers to long-distance quantum communication architectures such as quantum repeaters.

\section*{Code Availability}

The code used in this work is publicly available in the GitHub repository \href{https://github.com/mahsakarimii/perturbative-ZPG-code}{perturbative-ZPG-code}, which is also cited as Ref.~\cite{karimi_perturbative_zpg_code}.

\section*{Contributions}
S.W. and M.K. conceived the idea. M.K. performed the derivations, numerical simulations, and wrote the manuscript. S.M. provided feedback on the formalism and helped edit the paper. S.W. and C.S. co-supervised the project and helped edit the paper.

\section*{Acknowledgment}
This work is funded by the NSERC Alliance quantum consortia grants ARAQNE and QUINT and partially funded by the TUF-TOPIQC
project part of the Trilateral Call for Quantum Innovation co-financed by Germany, the Netherlands and
by the French National Quantum Strategy (France
2030) program. We would like to thank Arsalan Motamedi for the useful discussion.

\bibliography{ref}

\clearpage

\onecolumngrid
\appendix

\section{Technical details of our approach}\label{app:proofs}

\begin{figure}[h]
\centering
\begin{tikzpicture}[scale=1.5]

\draw[line width = 1.2] (0,2) -- (2,2);
\draw[line width = 1.2] (0,0) -- (2,0);

\node[left] at (0,2) {\( \boldsymbol{\ket{e_i}} \)};
\node[left] at (0,0) {\( \boldsymbol{\ket{g_i}} \)};

\draw[->, line width = 1.2, decayred] (1,2) -- (1,0.1);

\draw[->, decorate, line width = 1.2, decoration={snake, amplitude=1mm, segment length=5mm}, photonblue] (1,1.2) -- (2.5,1.2);
\node[photonblue] at (3.5,1.2) {\( \textbf{a photon at mode } \boldsymbol{i}\)};

\end{tikzpicture}
\caption{Each emission mode corresponds to a jump operator $\sigma_i = \ket{g_i}\bra{e_i}$. This effect is captured by anti-commutator terms in \cref{eq:zero-order-zpg}.}
\label{fig:sigma_i}
\end{figure}
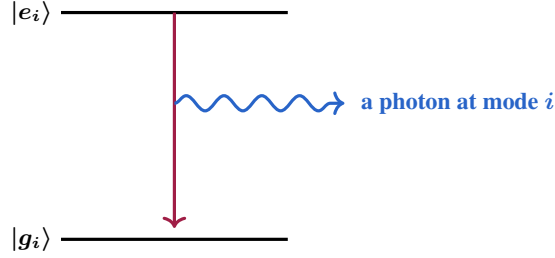

Here we provide the proofs of our perturbative expansions in \cref{sec:method}. Following the notation introduced in \cref{sec:method}, our ZPG Lindbladian can be decomposed as
\begin{align}
\mathcal L_{\mathrm{ZPG}} = \mathcal L_{\mathrm{ZPG}}^{\mathrm{ideal}} + \sum_k \gamma_k \mathcal L_k,
\end{align}
where the zero-order zero-photon generator is given by
\begin{align}\label{eq:zero-order-zpg}
\mathcal L_{\mathrm{ZPG}}^{\mathrm{ideal}}[\bullet] = -i[H, \bullet] - \frac12 \gamma \sum_i \{\sigma_i^\dagger \sigma_i, \bullet \},
\end{align}
where each $\sigma_i = \ket{g_i}\bra{e_i}$ corresponds to an emission in a particular mode. See \cref{fig:sigma_i} for a demonstration.

We note that \cref{eq:zero-order-zpg} can be re-written as
\begin{align}
\mathcal L_{\mathrm{ZPG}}^{\mathrm{ideal}}[\bullet] = -i( H_{\mathrm{eff}}\bullet - \bullet H_{\mathrm{eff}}^\dag),
\end{align}
with
\begin{align}
H_{\mathrm{eff}} = H - \frac{i}2 \gamma \sum_i \ket{e_i}\bra{e_i}.
\end{align}
Hence, assuming the Hamiltonian $H$ is time-independent, we obtain that
\begin{align}
\mathcal P_0^{(0)}(t,t_0)[\bullet] = e^{-iH_{\mathrm{eff}}(t-t_0)} \bullet e^{iH_{\mathrm{eff}}^\dagger(t-t_0)}.
\end{align}

\paragraph*{Assumptions} We use the following assumptions in simplifying the expressions, which help us get our perturbative expansions:
\begin{enumerate}
    \item We assume all ideal Hamiltonians are diagonal, with ground and excited states $(\ket{g_i},\ket{e_i})_i$, where a photon is emitted from the transition $\ket{e_i} \to \ket{g_i}$. We call each pair $\ket{g_i},\ket{e_i}$ an emission level-pair. We allow multiple transitions for each emitter.

    \item For each emission level-pair, we have the jump operator $\sigma_i:= \ket{g_i}\bra{e_i}$, and we assume $[\sigma_i,\sigma_j]=0$ for all $i,j$. This is satisfied so long as there is no cascade emissions i.e., $\ket{e_i}\neq\ket{g_j}$ for all emission level-pairs $i, j$. Our condition excludes systems such as biexciton quantum dots, but includes all single-photon sources.

    \item Lastly, we assume that all emission level-pair have the same energy gap. Note that if this condition is not satisfied, we can readily go to the rotating frame of the emitters and ensure this is the case.
\end{enumerate}

We will now proceed to calculate zero-order solutions to arbitrary post-selection maps.

\subsection{Zero-order solutions}

Here we apply the ZPG framework to the zero-order case (i.e., ideal evolutions). We assume all excitations have the same energy i.e., $\omega_{e_i} - \omega_{g_i} = \Delta$ for all $i$. If this is not the case, we can go to the rotating frame of the local emitters, making $\Delta=0$. We start from
\begin{align}\label{eq:commute-trick}
e^{-iH_{\mathrm{eff}}\tau} D_i e^{iH_{\mathrm{eff}}\tau} = e^{-\frac\gamma2\tau+ i\Delta \tau} D_i,
\end{align}
which would help us derive our zero-order solutions. 

We prove \cref{eq:zero-order-closed-form} inductively. We use $\mathbf m=0$ as our base of induction. Now, we move on to the induction step: assuming \cref{eq:zero-order-closed-form} holds for all $\mathbf n$ with $|\mathbf m| = K$, we prove if for all $|\mathbf m| = K+1$. Consider such a pattern $\mathbf m$. Recall the recursion relation
\begin{align}
\mathcal P_{\mathbf m}^{(0)} = \sum_{j=1}^N \int_{t_0}^t \mathcal P_0^{(0)}(t,t') \, \mathcal J_j \, \mathcal P^{(0)}_{\mathbf m - \mathbf e_j} (t',0)\, \mathrm dt'.
\end{align}
Plugging \cref{eq:zero-order-closed-form} in the above gives
\begin{align}\label{eq:U_m0-proof}
\mathcal P_{\mathbf m}^{(0)}[\bullet] = \sum_{j=1}^N \int_{t_0}^t f_{\mathbf m - \mathbf e_j}(t') A_j(t') \bullet A_j^\dagger(t'), 
\end{align}
where
\begin{align}
\begin{split}
A_j(t') &= e^{-iH_{\mathrm{eff}}(t-t')} D_j e^{-iH_{\mathrm{eff}}(t'-t_0)} \mathbf D^{\mathbf m - \mathbf e_j}\\
&= e^{-\frac\gamma2t' + i\Delta t'}e^{-iH_{\mathrm{eff}}(t-t_0)} \mathbf D^{\mathbf m},
\end{split}
\end{align}
where in the second line we have used \cref{eq:commute-trick}. Plugging back into \cref{eq:U_m0-proof} gives
\begin{align}
\begin{split}
\mathcal P_{\mathbf m}^{(0)}[\bullet] &= \left(\sum_{j=1}^N \int_{t_0}^t e^{-\gamma t'}f_{\mathbf m - \mathbf e_j}\right) e^{-iH_{\mathrm{eff}}(t-t_0)} \mathbf D^{\mathbf m}\bullet \mathbf D^{\mathbf m} {}^\dagger e^{iH^\dagger_{\mathrm{eff}}(t-t_0)}\\
&= f_{\mathbf m} e^{-iH_{\mathrm{eff}}(t-t_0)} \mathbf D^{\mathbf m}\bullet \mathbf D^{\mathbf m} {}^\dagger e^{iH^\dagger_{\mathrm{eff}}(t-t_0)},
\end{split}
\end{align}
where we have defined $f_{\mathbf m}$ recursively via
\begin{align}
\begin{split}
f_{\mathbf m}(t) &= \sum_{j=1}^N \int_{t'=0}^t e^{-\gamma t'} f_{\mathbf m - \mathbf e_j}(t') \, \mathrm dt',\\
\text{with} \quad f_{0}(t) &= 1.
\end{split}
\end{align}
This implies
\begin{align}
\begin{split}
f'_{\mathbf m}(t) &= e^{-\gamma t'} \sum_j f_{\mathbf m - \mathbf e_j}(t),\\
\text{with} \quad f_{\mathbf m}(0) &= 0\quad \forall \mathbf m \neq 0,\\
f_0(t) &= 1.
\end{split}
\end{align}
It is straightforward to verify that \cref{eq:f-solution} satisfies the above equations as
\begin{align}
\begin{split}
f'_{\mathbf m}(t) &= \frac{|\mathbf m| }{\mathbf m!} (1-e^{-\gamma t})^{|\mathbf m|-1} \cdot e^{-\gamma t}\\
&= e^{-\gamma t} \sum_j \frac{m_j}{\mathbf m!} (1-e^{-\gamma t})^{|\mathbf m| - 1}\\
&= e^{-\gamma t} \sum_{j} \frac{1}{(\mathbf m-\mathbf e_j)!} (1-e^{-\gamma t})^{|\mathbf m - \mathbf e_j|}\\
&= e^{-\gamma t'} \sum_j f_{\mathbf m - \mathbf e_j} (t).
\end{split}
\end{align}

\subsection{First-order solutions}
\label{app:1storder}
We follow a standard approach similar to our previous work \cite{karimi2026comparing} to compute the first-order corrections to zero-photon map $\mathcal P_0$. Subsequently, we use the recursion relation \cref{eq:ZPG} to calculate first-order corrections to other photon-counted maps.

Recall that the perturbed zero-photon generator is given by
\begin{align}
\mathcal L_{\mathrm{ZPG}} = \mathcal L_{\mathrm{ZPG}}^{\mathrm{ideal}} + \sum_k \gamma_k \mathcal L_k^{\mathrm{err}}.
\end{align}
Let $\rho(t)$ be a solution to
\begin{align}\label{eq:rho-of-t}
\dot\rho(t) = \mathcal L_{\mathrm{ZPG}}^{\mathrm{ideal}}[\rho] +\sum_k \gamma_k \mathcal L_{k}^{\mathrm{err}}[\rho]. 
\end{align}
Our goal is to find a perturbative expansion for $\rho$. Going to the interaction picture defined via $H_{\mathrm{eff}}$ we have
\begin{align}
\rho_I(t) = e^{iH_{\mathrm{eff}}(t-t_0)} \rho(t) e^{-iH_{\mathrm{eff}}^\dagger (t-t_0)}.
\end{align}
Plugging into \cref{eq:rho-of-t} gives
\begin{align}
\dot\rho_I(t) = \sum_{k}\gamma_k e^{iH_{\mathrm{eff}}(t-t_0)} \mathcal L_{k}\left[e^{-iH_{\mathrm{eff}}(t-t_0)}\rho_I(t) e^{-iH_{\mathrm{eff}}^\dagger(t-t_0)}\right] e^{-iH_{\mathrm{eff}}^\dagger(t-t_0)},
\end{align}
hence, setting $\mathcal L_{k,I}[\bullet]:= e^{iH_{\mathrm{eff}}(t-t_0)} \mathcal L_{k}^{\mathrm{err}}\left[e^{-iH_{\mathrm{eff}}(t-t_0)}\bullet e^{-iH_{\mathrm{eff}}^\dagger(t-t_0)}\right] e^{-iH_{\mathrm{eff}}^\dagger(t-t_0)}$ we get that $\rho_I$ is the solution to the following integral equation
\begin{align}
\rho_I(t) = \sum_k \gamma_k \int_{t_0}^t \mathcal L_{k,I}[\rho_I(t')]\, \mathrm dt'.
\end{align}
We can decompose $\rho_I(t)$ into a zero and first order parts, where the zero order solution is $\rho_I(t) = \rho_I(0) = \rho(0)$. We get the first-order solution by setting
\begin{align}
\rho_I^{(1)}(t) = \sum_k \gamma_k \int_{t_0}^t \mathcal L_{k,I}[\rho(0)]\, \mathrm dt',
\end{align}
which gives
\begin{align}
\rho^{(1)}(t) &= \sum_k \gamma_k \int_{t_0}^t e^{-iH_{\mathrm{eff}}(t-t')}\mathcal L_{k}\left[e^{-iH_{\mathrm{eff}}(t'-t_0)}\rho^{(0)}(t) e^{-iH_{\mathrm{eff}}^\dagger(t'-t_0)}\right] e^{iH_{\mathrm{eff}}^\dagger(t-t')}\, \mathrm dt'\\
&= \sum_{k}\gamma_k \int_{t_0}^t \mathcal P_0^{(0)}(t,t') \, \mathcal L_k^{}\, \mathcal P_{0}^{(0)}(t',t_0)[\rho^{(0)}(t)]\, \mathrm dt',
\end{align}
which immediately gives
\begin{align}\label{eq:base-for-induction}
\mathcal P^{(1)}_0 = \sum_k \gamma_k \int_{t_0}^t \mathcal P_0^{(0)}(t,t') \, \mathcal L_k^{\mathrm{err}}\, \mathcal P_0^{(0)}(t',t_0)\, \mathrm dt'.
\end{align}
In what follows, we use \cref{eq:base-for-induction} as a base for an induction to obtain \cref{eq:first-order-solution}.

It is straightforward to use \cref{eq:ZPG} to obtain
\begin{align}\label{eq:recursion-for-first-order}
\mathcal P_{\mathbf m}^{(1)}(t,t_0) = \sum_j \int_{t_0}^t \mathcal P_{0}^{(0)}(t,t') \mathcal J_j \,\mathcal P_{\mathbf m - \mathbf e_j}^{(1)}(t',t_0) \, \mathrm dt' + \sum_j \int_{t_0}^t \mathcal P_{0}^{(1)}(t,t') \mathcal J_j \,\mathcal P_{\mathbf m-\mathbf e_j}^{(0)}(t',t_0) \, \mathrm dt'.
\end{align}
Let us use $\mathcal L^{\mathrm{err}} := \sum_k \gamma_k \mathcal L_k$. Recall that we eventually want to prove
\begin{align}\label{eq:first-order-solution-app}
\mathcal P_{\mathbf m}^{(1)}(t,t_0) = \sum_{\mathbf n \leq \mathbf m} \int_{t'=t_0}^t \mathcal P_{\mathbf n}^{(0)}(t,t') \mathcal L^{\mathrm{err}} \mathcal P_{\mathbf m-\mathbf n}^{(0)}(t',t_0)\, \mathrm dt'.
\end{align}
In order to prove \cref{eq:first-order-solution-app} inductively, we assume the relation holds for all $\mathbf m\in\mathbb N^N$ with $|\mathbf m| = k-1$ and we prove it for $|\mathbf m| = k$. Plugging the induction assumption (i.e., the formula \cref{eq:first-order-solution-app} for all $\mathbf m - \mathbf e_j$) into \cref{eq:recursion-for-first-order} gives
\begin{align}\label{eq:pf-of-1st-order}
\begin{split}
\mathcal P_{\mathbf m}^{(1)}(t,t_0) &= \sum_j \sum_{\mathbf n\leq \mathbf m - \mathbf e_j} \int_{t'=t_0}^t\int_{t''=t_0}^{t'} \mathcal P_0^{(0)}(t,t') \mathcal J_j\, \mathcal P_{\mathbf n}^{(0)}(t',t'') \mathcal L^{\mathrm{err}} \mathcal P^{(0)}_{\mathbf m - \mathbf n - \mathbf e_j}(t'',t_0)\, \mathrm dt'' \mathrm dt'\\
&+ \sum_j \int_{t'=t_0}^t \int_{t''=t'}^t \mathcal P_0^{(0)}(t,t'') \mathcal L^{\mathrm{err}}\mathcal P_{0}^{(0)}(t'',t') \mathcal J_j \mathcal P_{\mathbf m - \mathbf n - \mathbf e_j}^{(0)}(t',t_0)\, \mathrm dt'' \mathrm dt',\\
&\overset{(a)}{=} \sum_j \sum_{\mathbf n\leq \mathbf m - \mathbf e_j} \int_{t'=t_0}^t\int_{t''=t_0}^{t'} \mathcal P_0^{(0)}(t,t') \mathcal J_j\, \mathcal P_{\mathbf n}^{(0)}(t',t'') \mathcal L^{\mathrm{err}} \mathcal P^{(0)}_{\mathbf m - \mathbf e_j}(t'',t_0)\, \mathrm dt'' \mathrm dt'\\
&+ \int_{t'=t_0}^t \mathcal P_0^{(0)} \mathcal L^{\mathrm{err}} \mathcal P_{\mathbf m}^{(0)}(t',t_0)\, \mathrm dt', 
\end{split}
\end{align}
where $(a)$ follows from rewriting the second term in the first equation of \cref{eq:pf-of-1st-order} as
\begin{align}
\begin{split}
&\sum_j \int_{t'=t_0}^t \int_{t''=t'}^t \mathcal P_0^{(0)}(t,t'') \mathcal L^{\mathrm{err}}\mathcal P_{0}^{(0)}(t'',t') \mathcal J_j \mathcal P_{\mathbf m - \mathbf e_j}^{(0)}(t',t_0)\, \mathrm dt'' \mathrm dt'\\
&= \int_{t''=t_0}^t  \mathcal P_{0}^{(0)}(t,t'') \mathcal L^{\mathrm{err}} \sum_j \int_{t'=t_0}^{t''} \mathcal P_0^{(0)}(t'',t') \mathcal J_j \mathcal P_{\mathbf m - \mathbf e_j}^{(0)}(t',0)\, \mathrm dt' \mathrm dt''\\
&= \int_{t''=t_0}^t \mathcal P_0^{(0)} \mathcal L^{\mathrm{err}} \mathcal P_{\mathbf m}^{(0)}\, \mathrm dt''.
\end{split}
\end{align}
where in the last line we have made use of \cref{eq:ZPG} again. We can also simplify the first term in \cref{eq:pf-of-1st-order} as follows
\begin{align}\label{eq:very-long}
\begin{split}
&\sum_j \sum_{\mathbf n\leq \mathbf m - \mathbf e_j} \int_{t'=t_0}^t\int_{t''=t_0}^{t'} \mathcal P_0^{(0)}(t,t') \mathcal J_j\, \mathcal P_{\mathbf n}^{(0)}(t',t'') \mathcal L^{\mathrm{err}} \mathcal P_{\mathbf m - \mathbf n - \mathbf e_j}^{(0)}(t'',t_0)\, \mathrm dt'' \mathrm dt'\\
&= \sum_j \sum_{\mathbf e_j \leq \mathbf n'\leq \mathbf m} \int_{t'=t_0}^t\int_{t''=t_0}^{t'} \mathcal P_0^{(0)} (t,t') \mathcal J_j \mathcal P_{\mathbf n'-\mathbf e_j}^{(0)}(t',t'') \mathcal L^{\mathrm{err}} \mathcal P_{\mathbf m - \mathbf n'}^{(0)}(t'',t_0)\, \mathrm dt''\mathrm dt'\\
&\overset{(b)}{=} \sum_j \sum_{\mathbf n'\leq \mathbf m} \int_{t'=t_0}^t\int_{t''=t_0}^{t'} \mathcal P_0^{(0)} (t,t') \mathcal J_j \mathcal P_{\mathbf n'-\mathbf e_j}^{(0)}(t',t'') \mathcal L^{\mathrm{err}} \mathcal P_{\mathbf m - \mathbf n'}^{(0)}(t'',t_0)\, \mathrm dt''\mathrm dt'\\
&= \sum_{\mathbf n'\leq \mathbf m} \int_{t''=0}^t \underbrace{\sum_j \int_{t'=t''}^{t_0} \mathcal P_{0}^{(0)}(t,t') \mathcal J_j \mathcal P_{\mathbf n' - \mathbf e_j}^{(0)}(t',t'')}_{\mathcal P^{(0)}_{\mathbf n'}(t,t'') \text{ if }\mathbf n'\neq 0,\quad \text{otherwise }0} \mathcal L^{\mathrm{err}} \mathcal P_{\mathbf m - \mathbf n'}^{(0)}(t'',t_0) \, \mathrm dt' \mathrm dt''\\
&= \sum_{0\neq \mathbf n'\leq \mathbf m} \int_{t''=0}^t \mathcal P_{\mathbf n'}^{(0)}(t,t'') \mathcal L^{\mathrm{err}} \mathcal P_{\mathbf m - \mathbf n'}^{(0)}(t'',t_0)\, \mathrm dt'',
\end{split}
\end{align}
where in $(b)$ we have used the fact that $\mathcal P_{\mathbf n'-\mathbf e_j}$ is by definition zero, whenever $\mathbf n'\ngeq \mathbf e_j$, and hence, we can remove the condition $\mathbf e_j \leq \mathbf n'$ from the summation conditions.
Lastly, plugging \cref{eq:very-long} into \cref{eq:pf-of-1st-order} gives
\begin{align}
\mathcal P_{\mathbf m}^{(1)}(t,t_0) = \sum_{\mathbf n \leq \mathbf m} \int_{t'=t_0}^t \mathcal P_{\mathbf n}^{(0)}(t,t') \mathcal L^{\mathrm{err}} \mathcal P_{\mathbf m-\mathbf n}^{(0)}(t',t_0)\, \mathrm dt',
\end{align}
completing our induction.

\subsection{Higher-order solutions}\label{sec:higher-order}

In this section, our aim is to generalize our formulas to higher-order errors. Later in this section, we prove that for the $k$-th order perturbation
\begin{align}\label{eq:order-k}
\begin{split}
\mathcal P_{\mathbf m}^{(k)}(t,t_0) &= \sum_{j_1,\cdots, j_k} \Pi_{i=1}^k \gamma_{j_i} \sum_{\sum_i{\mathbf n_i}= \mathbf m} \int_{t'_1=t_0}^{t} \int_{t'_2=t'_1}^{t}\cdots\int_{t'_{k}=t'_{k-1}}^t \mathcal P_{\mathbf n_{k+1}}^{(0)}(t,t'_k)\mathcal L_{j_k}^{\mathrm{err}} \mathcal P_{\mathbf n_{k}}^{(0)}(t'_k,t'_{k-1}) \\
&\qquad\qquad\qquad\qquad\qquad\qquad\qquad\qquad\qquad\qquad\cdots \mathcal P_{\mathbf n_2}^{(0)}(t'_2,t'_1)\mathcal L_{j_1}^{\mathrm{err}} \mathcal P_{\mathbf n_{1}}^{(0)}(t'_1,t_0)\, \mathrm dt'_k\cdots\mathrm dt'_1
\end{split}
\end{align}
Let us provide a physical intuition for \cref{eq:order-k}. The $k$-th order error is the summation over the possibilities of the following: having emitted $\mathbf n_1$ photons, then incurring one of the Lindbladian errors, subsequently measuring $\mathbf n_2$ photons and then incurring another Lindbladian error, and so on, until emitting $\mathbf n_{k+1}$ photons at the end. We have the condition that the total number of emitted photons must match $\mathbf m$, that is $\mathbf n_1 + \cdots + \mathbf n_{k+1} = \mathbf m$.

As a special case, we have that second-order errors are given by
\begin{align}
\mathcal P_{\mathbf m}^{(2)}(t,t_0) = \sum_{k,j} \gamma_k \gamma_j \sum_{\mathbf n_1+
\mathbf n_2\leq \mathbf m} \int_{t'=t_0}^t \int_{t''=t_0}^{t'} \mathcal P_{\mathbf n_1}^{(0)}(t,t') \mathcal L_k^{\mathrm{err}}\mathcal P_{\mathbf n_2}^{(0)}(t',t'') \mathcal L_{j}^{\mathrm{err}} \mathcal P_{\mathbf m - \mathbf n_1 - \mathbf n_2}^{(0)}(t'',t_0)\, \mathrm dt'.
\end{align}
This is the formula used in the main text. 

In what follows, we prove the above formulas.

\subsubsection{Proof of the higher-order formulas}

Here, we prove the general $k$-th order perturbation formula \cref{eq:order-k} using a recursive approach. To avoid tedious calculations, our proof here follows a less direct approach than the previous section. We start by defining a generating map for photon-resolved propagators: We first introduce formal variables $\mathbf z=(z_1,\dots,z_N)$ and define the multivariate generating superoperator
\begin{align}\label{eq:gen-def}
\mathcal P(\mathbf z;t,t_0)
:=\sum_{\mathbf m\in\mathbb N^N}\mathbf z^{\mathbf m}\,\mathcal P_{\mathbf m}(t,t_0),
\qquad 
\mathbf z^{\mathbf m}:=\prod_{j=1}^N z_j^{m_j}.
\end{align}
We also define the $\mathbf z$-dependent generator
\begin{align}\label{eq:gen-generator}
\mathcal G(\mathbf z):=\mathcal L_{\mathrm{ZPG}}+\sum_{j=1}^N z_j\,\mathcal J_j,
\end{align}
so that $\mathcal P(\mathbf z;t,t_0)$ satisfies the differential equation.
\begin{align}\label{eq:gen-ode}
\partial_t \mathcal P(\mathbf z;t,t_0)=\mathcal G(\mathbf z)\,\mathcal P(\mathbf z;t,t_0),
\qquad \mathcal P(\mathbf z;t_0,t_0)=\mathcal I,
\end{align}
which is derived by differentiating \cref{eq:gen-def}, and using the ZPG recursive formula \cref{eq:ZPG}. Indeed, we can extract the coefficient of $\mathbf z^{\mathbf m}$ on both sides of \cref{eq:gen-ode} to obtain
\begin{align}\label{eq:hierarchy}
\partial_t \mathcal P_{\mathbf m}(t,t_0)
=\mathcal L_{\mathrm{ZPG}}\mathcal P_{\mathbf m}(t,t_0)
+\sum_{j=1}^N \mathcal J_j\,\mathcal P_{\mathbf m-\mathbf e_j}(t,t_0),
\end{align}
with the convention $\mathcal P_{\mathbf r}\equiv 0$ whenever $\mathbf r\notin\mathbb N^N$. 
Note that \cref{eq:hierarchy} is the differential form of the ZPG recursion relation \cref{eq:ZPG}. Next, decompose
\begin{align}
\mathcal G(\mathbf z)=\underbrace{\Big(\mathcal L_{\mathrm{ZPG}}^{\mathrm{ideal}}+\sum_{j=1}^N z_j\,\mathcal J_j\Big)}_{=:~\mathcal G^{(0)}(\mathbf z)}
~+~\mathcal L^{\mathrm{err}}.
\end{align}
Let $\mathcal P^{(0)}(\mathbf z;t,t_0)$ denote the propagator generated by $\mathcal G^{(0)}(\mathbf z)$ (i.e., the solution of \cref{eq:gen-ode} with $\mathcal L^{\mathrm{err}}=0$). By construction,
\begin{align}\label{eq:gen-coeff-zero}
\mathcal P^{(0)}(\mathbf z;t,t_0)=\sum_{\mathbf m\in\mathbb N^N}\mathbf z^{\mathbf m}\,\mathcal P_{\mathbf m}^{(0)}(t,t_0).
\end{align}
Here we make use of an auxillary fact: Consider linear maps $X, A, B$ (can be matrices, or super-operators). In this line of reasoning, we use the basic fact that if $\dot X(t) = A(t) X(t) + B(t) X(t)$, with auxiliary variable $U_A$ defined via $\partial_t U_A(t,s) = A(t) U_A(t,s)$ with $U(s,s) = \mathbb I$. We then have $X(t) = U_A(t,0) X(0) + \int_{0}^t U_A(t,\tau) B(\tau) X(\tau)\, \mathrm d\tau$.

\begin{align}\label{eq:duhamel}
\mathcal P(\mathbf z;t,t_0)
=\mathcal P^{(0)}(\mathbf z;t,t_0)
+\int_{t_0}^{t}\mathcal P^{(0)}(\mathbf z;t,\tau)\,\mathcal L^{\mathrm{err}}\,\mathcal P(\mathbf z;\tau,t_0)\,\mathrm d\tau.
\end{align}
Expanding $\mathcal P(\mathbf z)=\sum_{k\ge 0}\mathcal P^{(k)}(\mathbf z)$ into homogeneous orders in the perturbation parameters $\{\gamma_j\}$ and matching the $k$-th order terms yields, for $k\ge 1$,
\begin{align}\label{eq:order-recursion-gen}
\mathcal P^{(k)}(\mathbf z;t,t_0)
=\int_{t_0}^{t}\mathcal P^{(0)}(\mathbf z;t,\tau)\,\mathcal L^{\mathrm{err}}\,\mathcal P^{(k-1)}(\mathbf z;\tau,t_0)\,\mathrm d\tau.
\end{align}
Taking the coefficient of $\mathbf z^{\mathbf m}$ on both sides of \cref{eq:order-recursion-gen} and using the Cauchy-product convolution implied by \cref{eq:gen-coeff-zero}, we obtain the following recursion for the photon-resolved maps:
\begin{align}\label{eq:order-recursion-m}
\mathcal P_{\mathbf m}^{(k)}(t,t_0)
=\sum_{\mathbf n\le \mathbf m}\int_{t_0}^{t}
\mathcal P_{\mathbf n}^{(0)}(t,\tau)\,\mathcal L^{\mathrm{err}}\,\mathcal P_{\mathbf m-\mathbf n}^{(k-1)}(\tau,t_0)\,\mathrm d\tau.
\end{align}
For $k=1$, \cref{eq:order-recursion-m} reduces to \cref{eq:first-order-solution-app}.

We now prove \cref{eq:order-k} by induction on $k$ using \cref{eq:order-recursion-m}. For clarity, we present an equivalent form of \cref{eq:order-k} in which the photon counts are explicitly partitioned into $k+1$ segments:
\begin{align}\label{eq:order-k-partition}
\begin{split}
\mathcal P_{\mathbf m}^{(k)}(t,t_0)
&=\sum_{j_1,\dots,j_k}\Big(\prod_{r=1}^k\gamma_{j_r}\Big)
\sum_{\substack{\mathbf n_1,\dots,\mathbf n_{k+1}\in\mathbb N^N\\ \mathbf n_1+\cdots+\mathbf n_{k+1}=\mathbf m}}
\int_{t_0\le t'_1\le \cdots \le t'_k\le t}
\mathcal P_{\mathbf n_{k+1}}^{(0)}(t,t'_k)\,\mathcal L_{j_k}\,
\mathcal P_{\mathbf n_{k}}^{(0)}(t'_k,t'_{k-1}) \cdots \\
&\hspace{3.8cm}\cdots\, \mathcal L_{j_1}\,\mathcal P_{\mathbf n_{1}}^{(0)}(t'_1,t_0)\,
\mathrm dt'_1\cdots \mathrm dt'_k.
\end{split}
\end{align}
Equation \cref{eq:order-k-partition} is equivalent to \cref{eq:order-k} by setting
$\mathbf n_{k+1}=\mathbf m-\sum_{r=1}^{k}\mathbf n_r$ and rewriting the sum accordingly.\\

\textit{Proof of \cref{eq:order-k-partition}}:
The base case $k=0$ is immediate, as the simplex integral is empty and the only decomposition is $\mathbf n_1=\mathbf m$, yielding $\mathcal P_{\mathbf m}^{(0)}(t,t_0)$.

For the inductive step, assume \cref{eq:order-k-partition} holds for $k-1$. Starting from \cref{eq:order-recursion-m}, we write
\begin{align}
\begin{split}
\mathcal P_{\mathbf m}^{(k)}(t,t_0)
&=\sum_{\mathbf n_{k+1}\le \mathbf m}\int_{t_0}^{t}
\mathcal P_{\mathbf n_{k+1}}^{(0)}(t,t'_k)\,\mathcal L^{\mathrm{err}}\,\mathcal P_{\mathbf m-\mathbf n_{k+1}}^{(k-1)}(t'_k,t_0)\,\mathrm dt'_k\\
&=\sum_{j_k}\gamma_{j_k}\sum_{\mathbf n_{k+1}\le \mathbf m}\int_{t_0}^{t}
\mathcal P_{\mathbf n_{k+1}}^{(0)}(t,t'_k)\,\mathcal L_{j_k}^{\mathrm{err}}\,\mathcal P_{\mathbf m-\mathbf n_{k+1}}^{(k-1)}(t'_k,t_0)\,\mathrm dt'_k.
\end{split}
\end{align}
Applying the induction hypothesis to $\mathcal P_{\mathbf m-\mathbf n_{k+1}}^{(k-1)}(t'_k,t_0)$ (with final time $t'_k$ instead of $t$) yields
\begin{align}
\begin{split}
\mathcal P_{\mathbf m-\mathbf n_{k+1}}^{(k-1)}(t'_k,t_0)
&=\sum_{j_1,\dots,j_{k-1}}\Big(\prod_{r=1}^{k-1}\gamma_{j_r}\Big)
\sum_{\substack{\mathbf n_1,\dots,\mathbf n_{k}\in\mathbb N^N\\ \mathbf n_1+\cdots+\mathbf n_{k}=\mathbf m-\mathbf n_{k+1}}}
\int_{t_0\le t'_1\le \cdots \le t'_{k-1}\le t'_k}
\mathcal P_{\mathbf n_{k}}^{(0)}(t'_k,t'_{k-1})\,\mathcal L_{j_{k-1}}^{\mathrm{err}}\cdots \\
&\hspace{3.8cm}\cdots\,\mathcal L_{j_1}^{\mathrm{err}}\,\mathcal P_{\mathbf n_1}^{(0)}(t'_1,t_0)\,
\mathrm dt'_1\cdots \mathrm dt'_{k-1}.
\end{split}
\end{align}
Substituting this expression back and relabeling the summed indices gives a sum over
$j_1,\dots,j_k$ with prefactor $\prod_{r=1}^k\gamma_{j_r}$, as well as a sum over
$\mathbf n_1,\dots,\mathbf n_{k+1}$ satisfying
$\mathbf n_1+\cdots+\mathbf n_{k+1}=\mathbf m$. The time-ordering region becomes
$t_0\le t'_1\le\cdots\le t'_{k-1}\le t'_k\le t$, producing exactly \cref{eq:order-k-partition}.
This completes the induction and proves \cref{eq:order-k-partition}, and hence \cref{eq:order-k}.

\subsection{Detector inefficiency}\label{sec:eta}

The detector efficiency $\eta$ is often far from perfect and cannot be treated perturbatively. However, as there are only a few photons generated in light--matter interaction-based schemes, we can use convolution ideas similar to \cite{wein2020analyzing} to capture detector inefficiencies.
In other words, to account for loss, we take a classical distribution over the ideal solutions. For example, the state after detecting 2 overall photons with loss will involve all solutions corresponding to detecting $n\geq 2$. However, for the schemes we are interested in (i.e. those taking post-selected measurements that are rather robust to loss), the probability of producing more than two photons is small. Hence, with a finite summation (i.e., a mixture of perfect solutions), we can simulate the effect of loss as well. Concretely, letting $\tilde{\mathcal P}_{\mathbf m}$ and $\mathcal P_{\mathbf m}$ denote the conditional maps obtained by lossy and ideal detectors respectively, we have
\begin{align}\label{eq:eta}
\tilde{\mathcal P}_{\mathbf m} = \sum_{\mathbf n \geq \mathbf m} \eta^{|\mathbf m|} (1-\eta)^{|\mathbf n - \mathbf m|}\prod_{j=1}^N {n_j\choose m_j} \mathcal P_{\mathbf n}.
\end{align}
If we know our scheme can create at most $C$ photons  (e.g., for a RUS gate $C=2$), then we can restrict the summation to all $\mathbf n$ such that $|\mathbf n| \leq C$.

\section{General treatment of fidelity and Pauli error weights}\label{app:general-fidelity}

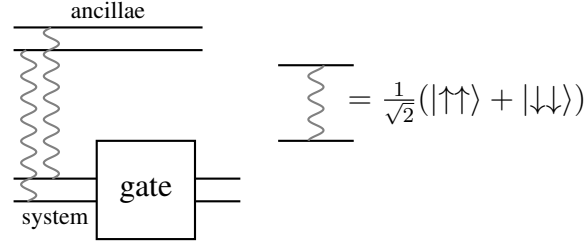
\begin{figure}
    \centering
\begin{tikzpicture}
\draw[thick, black] (-1,0) -- (1.5,0) node[midway, above] {\small ancillae};
\draw[thick, black] (-1,-.3) -- (1.5,-.3);
\draw[thick, black] (-1,-2) -- (0.1,-2);
\draw[thick, black] (-1,-2.3) -- (.1, -2.3)  node[midway, below] {system};
\draw[thick, black] (0.1, -1.5) rectangle (1.4, -2.8);
\node at (.75, -2.15) {\large gate};
\draw[thick, black] (1.4, -2) -- (2, -2);
\draw[thick, black] (1.4, -2.3) -- (2, -2.3);
\draw[thick, smooth, domain=-2:0, samples=100, color = gray] 
    plot ({-.5+0.1*sin(400*pi*\x)}, \x);
\draw[thick, smooth, domain=-2.3:-0.3, samples=100, color = gray] 
    plot ({-.8+0.1*sin(400*pi*\x)}, \x);

\draw[thick, black] (2.5,-.5) -- (3.5,-0.5);
\draw[thick, black] (2.5,-1.5) -- (3.5,-1.5);
\draw[thick, smooth, domain=-0.5:-1.5, samples=100, color = gray] 
    plot ({3+0.1*sin(400*pi*\x)}, \x);
\node at (5, -1) {\large $= \frac{1}{\sqrt2}(\ket{\uparrow\uparrow}+\ket{\downarrow\downarrow})$};
\end{tikzpicture}
\caption{The circuit computing the Choi matrix. The ancillae are $2$ dimensional registers and that the wavy lines represent the Bell state $\ket{\phi^+} = \frac1{\sqrt2}(\ket{\uparrow\uparrow}+\ket{\downarrow\downarrow})$. The entanglement fidelity ($F_{\mathrm{ent}}$) of a gate is the fidelity between the Choi matrices of the ideal channel and its implementation (i.e., \cref{eq:ent-fid}). Moreover, one can readily translate entanglement fidelity to average fidelity via \cref{eq:ent-to-avg}. The picture is taken from our previous work \cite{karimi2026comparing}.}
\label{fig:avg-fidelity-pic}
\end{figure}

Let the dimension of the system be $d$ (e.g., for $M$ emitters with local dimension $2$ we have $d=2^M$). We will show that quantities of the form 
\begin{align}\label{eq:general-overlap}
\bra{v}  (\mathcal V\otimes \mathcal P_{\mathbf m}) [\ket{\psi_{\mathrm{init}}}\bra{\psi_{\mathrm{init}}}]\ket{u},
\end{align}
can capture many interesting properties such as state and gate fidelities, as well as Pauli errors and the process ($\chi$) matrix. We replace $\mathcal P_{\mathbf m}$ with its perturbative expansion. We assume an initial pure $\rho_{\mathrm{init}}=\ket{\psi_{\mathrm{init}}}\bra{\psi_{\mathrm{init}}}$, and a unitary channel $\mathcal V$ which acts on a space of dimension at most $d$. 

Let us consider the `state fidelity' first. For a particular input state, say $\ket{\psi_{\mathrm{in}}}$, to the target state that we achieve by post-selecting on PNRD measurement $\mathbf m$. We denote this target state by $\ket{\psi_{\mathbf m}}$. Hence, the quantity of interest is
\begin{align}
F^{\mathrm{state}}_{\mathbf m} = \frac{1}{p_{\mathbf m}}\bra{\psi_{\mathbf m}} \mathcal P_{\mathbf m}[\ket{\psi_{\mathrm{in}}}\bra{\psi_{\mathrm{in}}}] \ket{\psi_{\mathbf m}},
\end{align}
with $p_{\mathbf m} = \mathrm{tr}(\mathcal P_{\mathbf m}[\ket{\psi_{\mathrm{in}}}\bra{\psi_{\mathrm{in}}}])$ being the success probability. We also use $p_{\mathbf m}^{(0)}$ to denote the success probability in the ideal (no noise) case.

Moving on to measures of gate fidelities, we explain how quantities such as the `average gate fidelity, `entanglement fidelity,' and `Pauli error weights' (or more generally the process matrix of the error channel) can be calculated via our approach. The Choi matrix associated to the photon-counted maps can be obtained via \cite{watrous2018theory}
\begin{align}
J_{\mathbf m}:= (\mathcal I \otimes \mathcal P_{\mathbf m})[\kket{\mathbb I}\bbra{\mathbb I}].
\end{align}
We also use $\ket{\phi^+}:=\frac{1}{\sqrt d}\kket{\mathbb I}$ to denote the maximally entangled state over two systems of local dimension $d$ (implying that $\mathbb I$ has size $d\times d$).

In quantum computation, we are interested in gate fidelities. The one that is well-defined in our case is the entanglement fidelity (which in deterministic cases is proportional to the average gate fidelity). The entanglement fidelity is defined as
\begin{align}\label{eq:ent-fid}
F_{\mathrm{ent}} := \bra{\phi^+} \left(\mathcal G^{-1}
\circ\tilde{\mathcal G}\left[\ket{\phi^+}\bra{\phi^+}\right]\right) \ket{\phi^+},
\end{align}
with $\phi^+$ being the maximally entangled state between our system and the environment, $\mathcal G$ is the ideal gate, and $\tilde{\mathcal G}$ is the noisy gate. In the case where $\tilde{\mathcal G}$ corresponds to the probabilistic application of a completely positive trace preserving (CPTP) map, the entanglement fidelity and the average fidelity are related via
\begin{align}\label{eq:ent-to-avg}
F_{\mathrm{avg}} = \frac{d F_{\mathrm{ent}} + 1}{d+1} ,
\end{align}
with $d$ being the dimension of our system \cite{nielsen2002simple}. Hence, our perturbative expansion allows for the computation of the entanglement fidelity.

Finally, we show that we can also compute the Pauli error weights. Note that letting $(\chi_{P,Q})_{P,Q}$ represent the $\chi$ matrix of our noisy channel, we have that
\begin{align}\label{eq:choi}
\chi_{P,Q} = \frac{1}{2^{2n}} \bbra{P} J_{\mathcal G}\kket{Q}.
\end{align}
where $\kket{P}$ and $\kket{Q}$ are vectorizations of Pauli operators. Again, as our approach allows us to readily compute the Choi matrix, we can compute the Pauli error weights as well.

\subsection{Fidelity expansion to first order}\label{app:fidelity-expansion}

Ref. \cite{karimi2026comparing} has achieved simple and easy-to-calculate expressions for infidelity. Inspired by that, we simplify the fidelity formula that one gets from our framework. 

Let us consider a target conditional state $\ket{\psi_{\mathbf m}}$ with error Lindbladian
\begin{align}\label{eq:error-lindbladian-app}
\mathcal L^{\mathrm{err}} =  \varepsilon \mathcal D[L].
\end{align}
For brievety we define jump operators
\begin{align}
K_{\mathbf r}(t_2,t_1):= \sqrt{f_{\mathbf r}(t_2-t_1)} e^{-iH_{\mathrm{eff}}(t_2-t_1)} \mathbf D^{\mathbf r}.
\end{align}
With this notation, we have
\begin{align}
\ket{\psi_{\mathbf m}} = \frac1{\sqrt{p_{\mathbf m}^{(0)}}} K_{\mathbf m}\ket{\psi_{\mathrm{init}}},
\end{align}
where $p^{(0)}_{\mathbf m}$ is the normalization factor, and it indicates the success probability in obtaining the outcome $\mathbf m$.

We are interested in computing the quantity
\begin{align}
F_{\mathbf m} = \frac{1}{p_{\mathbf m}}\langle \psi_{\mathbf m}| \rho_{\mathbf m}\ket{\psi_{\mathbf m}},
\end{align}
where $\rho_{\mathbf m}$ is the solution to the noisy ZPG equation, conditioned on measurement $\mathbf m$. Expanding to the first order, we write $p_{\mathbf m} = p_{\mathbf m}^{(0)} +\varepsilon p_{\mathbf m}^{(1)} + O(\varepsilon^2)$ and $\rho_{\mathbf m} = \rho_{\mathbf m}^{(0)} + \rho_{\mathbf m}^{(1)}+O(\varepsilon^2)$, which results in
\begin{align}
F_{\mathbf m} = 1 - \frac{\varepsilon}{p_{\mathbf m}^{(0)}} (p_{\mathbf m}^{(1)} -\bra{\psi_{\mathbf m}}\rho_{\mathbf m}^{(1)}\ket{\psi_{\mathbf m}}) + O(\varepsilon^2).
\end{align}
Since $p_{\mathbf m}^{(1)} = \mathrm{Tr}(\rho_{\mathbf m}^{(1)})$, we get
\begin{align}
F_{\mathbf m} = 1-\frac{\varepsilon}{p_{\mathbf m}^{(0)}} \mathrm{Tr}(\Pi_{\mathbf m}^{\perp} \rho_{\mathbf m}^{(1)}) + O(\varepsilon^2),
\end{align}
where $\Pi_{\mathbf m}^\perp := \mathbb I - \ket{\psi_{\mathbf m}}\bra{\psi_{\mathbf m}}$ is the projection onto the subspace orthogonal to the target state. Using our first-order solution \cref{eq:first-order-solution}, we get
\begin{align}
F_{\mathbf m} = 1-\frac{\varepsilon}{p_{\mathbf m}^{(0)}}\sum_{\mathbf n\le \mathbf m} \int_{t'=0}^t \mathrm{Tr}\left(\Pi_{\mathbf m}^\perp K_{\mathbf n}(t,t') \mathcal L^{\mathrm{err}} \left(K_{\mathbf m-\mathbf n}(t',0)\ket{\psi_{\mathrm{init}}}\bra{\psi_{\mathrm{init}}}K_{\mathbf m-\mathbf n}^\dag(t',0)\right) K_{\mathbf n}(t,t')^\dag \right)\, \mathrm dt' + O(\varepsilon^2).
\end{align}
Let us define $\ket{\phi_{\mathbf n}(t_2-t_1)}:=K_{\mathbf n}(t_2,t_1)\ket{\psi_{\mathrm{init}}}$. This allows us to write
\begin{align}
F_{\mathbf m} = 1-\frac{\varepsilon}{p_{\mathbf m}^{(0)}}\sum_{\mathbf n \le \mathbf m} \int_{t'=0}^t \mathrm{Tr}\left( \Pi_{\mathbf m}^\perp K_{\mathbf n}(t,t') \mathcal L^{\mathrm{err}} \left(\ket{\phi_{\mathbf m - \mathbf n}(t')}\bra{\phi_{\mathbf m -\mathbf n}(t')}\right) K_{\mathbf n}(t,t')^\dag \right)\, \mathrm dt'+O(\varepsilon^2).
\end{align}
We then have $K_{\boldsymbol\ell}(t_2,t_1)\ket{\phi_{\mathbf n}}\propto \ket{\phi_{\mathbf n + \boldsymbol\ell}}$, which due to multiplication by $\Pi_{\mathbf m}^\perp$, makes the $\{L^\dag L, \bullet\}$ term in $\mathcal L^{\mathrm{err}}$ cancel out (note that $\Pi_{\mathbf m}^\perp K_{\mathbf n} \ket{\phi_{\mathbf m-\mathbf n}}=0$). Therefore, after plugging \cref{eq:error-lindbladian-app} gives
\begin{align}
F_{\mathbf m} = 1-\frac{\varepsilon}{p_{\mathbf m}^{(0)}} \sum_{\mathbf n\le \mathbf m} \int_{t'=0}^t \left\|\Pi_{\mathbf m}^\perp K_{\mathbf n}(t,t') L K_{\mathbf m-\mathbf n}(t',0) \ket{\psi_{\mathrm{init}}}\right\|^2 + O(\varepsilon^2).
\end{align}

We highlight that a similar expression can be obtained for $p_{\mathbf m}$ in a perturbative way using $p_{\mathbf m} = \mathrm{Tr}(\rho_{\mathbf m}^{(1)})$ once again. Doing so, we obtain the form 
\begin{align}
p_{\mathbf m} = p_{\mathbf m}^{(0)} + \varepsilon \sum_{\mathbf n\le\mathbf m} \int_{0}^t \left( \|\ket{\phi_{L,\mathbf n}(t')}\|^2 - \mathrm{Re}\left( \langle \phi_{\mathbf n}(t')|\phi_{LL,\mathbf n}(t')\rangle \right) \right)\,\mathrm dt',
\end{align}
where
\begin{align}\label{eq:pert-in-prob}
\begin{split}
\ket{\phi_{\mathbf n}(t')} &= K_{\mathbf n}(t,t') K_{\mathbf m-\mathbf n}(t',0) \ket{\psi_{\mathrm{init}}},\\
\ket{\phi_{L,\mathbf n}(t')} &= K_{\mathbf n}(t,t') L K_{\mathbf m-\mathbf n}(t',0)\ket{\psi_{\mathrm{init}}},\\
\ket{\phi_{LL,\mathbf n}(t')} &= K_{\mathbf n}(t,t') L^\dag L K_{\mathbf m-\mathbf n}(t',0) \ket{\psi_{\mathrm{init}}}.
\end{split}
\end{align}

\paragraph*{Multiple Lindbladian error terms} Finally, we note that if the error Lindbladian is the sum of many terms, i.e., 
\begin{align}
\mathcal L^{\mathrm{err}} = \sum_i \gamma_i \mathcal D[L_i],
\end{align}
the first-order errors add linearly and we obtain
\begin{align}\label{eq:fidelity-formula}
F_{\mathbf m} = 1-\frac{1}{p_{\mathbf m}^{(0)}} \sum_{i} \gamma_i \sum_{\mathbf n\le \mathbf m} \int_{t'=0}^t \left\|\Pi_{\mathbf m}^\perp K_{\mathbf n}(t,t') L_i K_{\mathbf m-\mathbf n}(t',0) \ket{\psi_{\mathrm{init}}}\right\|^2 + O\left( (\sum_{i}\gamma_i)^2 \right).
\end{align}

One can interpret the integrand in \cref{eq:fidelity-formula} as the probability that the error jump terms ($L_i$) send us to the space orthogonal to $\ket{\psi_{\mathbf m}}$.

\subsection{Pauli error expansion to first order}\label{sec:pauli}

In this section, we achieve an expansion for Pauli error weights corresponding to a noisy photon-heralded operation.

Let $A \subset \mathbb N^N$ be a subset of measurement patterns that correspond to a successful implementation of our gate. We use $p_A = \sum_{\mathbf m} p_{\mathbf m}$ to denote the success probability of achieving a successful outcome. Also, let $R_{\mathbf m}$ denote the inverse of the desired gate together with corrections one needs to apply to achieve the right gate. For instance, \cref{fig:RUS-levels-and-table} lists these patterns along with their corrections for achieving $\mathsf{CZ}$, and as a particular example, we have $R_{(1010)}=\mathsf{CZ}^\dagger (S^\dagger\otimes S)$. Also, note that $R_{\mathbf m}$ is defined only on the computational basis (we extend its rows to the full Hilbert space by adding a zero block, in other words $R_{\mathbf m}\ket{e}$=0 for any state $\ket{e}$ not in the computational space, which is often the ground energy hyperfine levels). With that definition at hand, we note that the Choi matrix corresponding to the error channel corresponding to our implementation is
\begin{align}\label{eq:}
J^{\mathrm{err}} = \frac{1}{p_A} \sum_{\mathbf m\in A} (\mathbb I \otimes R_{\mathbf m}) J_{\mathbf m}(\mathbb I \otimes R_{\mathbf m}),
\end{align}
where $J_{\mathbf m}$ are Choi matrices corresponding to conditional maps $\mathcal P_{\mathbf m}$:
\begin{align}
J_{\mathbf m} = (\mathcal I \otimes \mathcal P_{\mathbf m})\left( \kket{\mathbb I}\bbra{\mathbb I} \right).
\end{align}
We recall that $\chi_{P,P} = \frac1{2^{2n}}\bbra{P} J^{\mathrm{err}}\kket{P}$.
We note that
\begin{align}\label{eq:blah}
R_{\mathbf m}\mathcal P_{\mathbf n}^{(0)}(t,t')\left[ L_j K_{\mathbf m -\mathbf n}(t',0) \rho K_{\mathbf m-\mathbf n}(t',0) L_j^\dag \right]R_{\mathbf m}^\dag = \tilde K_{\mathbf m,\mathbf n, j}\rho \tilde K_{\mathbf m,\mathbf n, j}^\dag,
\end{align}
basically applying a single Kraus term with
\begin{align}
\tilde K_{\mathbf m,\mathbf n, j} = R_{\mathbf m} K_{\mathbf n}(t,t') L_j K_{\mathbf m-\mathbf n}(t',0),
\end{align}
which would later help us simplify the contribution of $L_k \bullet L_k^\dag$ in expanding .On the other hand, note that
\begin{align}
\bbra{P} R_{\mathbf m} K_{\mathbf n}(t,t') K_{\mathbf m-\mathbf n}(t',0)\kket{\mathbb I} = 0,
\end{align}
for all $P\neq \mathbb I$, as $R_{\mathbf m} K_{\mathbf n}K_{\mathbf m - \mathbf n} \propto \mathbb I$. As a result, we obtain that the contribution of the terms $\{L_k^\dag L_k, \bullet\}$ to the Pauli error's expansion vanishes. Note that for the expansion of non-identity terms, we do not need to worry about $p_{A}^{(1)}$ as the numerator is already a first-order term (with no zero-order contribution).

Finally, using \cref{eq:blah}, we find
\begin{align}
\begin{split}
\chi_{P,P} 
&= \frac{1}{2^{2n} p_A^{(0)}} \sum_{\mathbf m\in A} \sum_{\mathbf n\le \mathbf m} \sum_j \gamma_j \int_{t'=0}^t \|\bbra{P} \tilde K_{\mathbf m, \mathbf n, j} \kket{\mathbb I}\|^2\, \mathrm dt'\\
&= \frac{1}{2^{2n} p_A^{(0)}} \sum_{\mathbf m\in A} \sum_{\mathbf n\le \mathbf m} \sum_j \gamma_j \int_{t'=0}^t \left|  \mathrm{Tr}\left( P\, R_{\mathbf m} K_{\mathbf n}(t,t') L_j K_{\mathbf m-\mathbf n}(t',0) \right) \right|^2\, \mathrm dt',
\end{split}
\end{align}
for any Pauli error $P\neq \mathbb I$.

\section{Application of the method in analyzing entanglement generation protocols: A case study based on single-photon heralding}\label{sec:n-protocol}

\begin{figure}[t]
\centering
\includegraphics[width=0.5\linewidth]{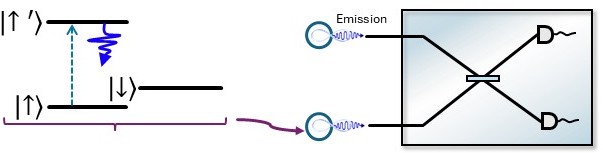}
\caption{In a single-photon heralded entanglement-generation protocol, each quantum emitter is modelled as a $\Lambda$-type system with two long-lived ground states $\ket{\uparrow}$ and $\ket{\downarrow}$ coupled to an excited state $\ket{\uparrow'}$. The emitter is first prepared in a superposition of the two ground states, after which a resonant laser pulse drives the $\ket{\uparrow}\!\rightarrow\!\ket{\uparrow'}$ transition. Spontaneous emission then entangles the internal state of the emitter with the photonic mode, producing a state of the form $(\ket{\uparrow1}+\ket{\downarrow0})/\sqrt{2}$, where $0$ and $1$ denote the photon number. Photons emitted from two distant emitters are interfered on a beam splitter placed midway between them, and a single-photon detection event projects (heralds) the two emitters into an entangled Bell state.
} 
\label{fig:n-protocol-levels}
\end{figure}

In this section, we discuss the $\mathsf{N}$ protocol from \cite{cabrillo1999creation,wein2020analyzing}, and analyze the effect of optical dephasing on this process. The $\mathsf N$ protocol is a standard single-photon heralding protocol for generating remote spin--spin entanglement. As depicted in \cref{fig:n-protocol-levels}, each of the two spatially separated $\Lambda$-type emitters, each with ground states $\ket{\uparrow}, \ket{\downarrow}$ and an optically excited state $\ket{\uparrow'}$, are initially prepared in $\ket{+} = (\ket{\uparrow} + \ket{\downarrow})/\sqrt{2}$. A resonant $\pi$-pulse on the $\ket{\uparrow} \leftrightarrow \ket{\uparrow'}$ transition maps this to $(\ket{\downarrow} + \ket{\uparrow'})/\sqrt{2}$ for each emitter. Subsequent spontaneous emission produces a single photon from each system, which are interfered on a 50{:}50 beam splitter located midway between the nodes.

Conditioned on the photonic detection pattern at the beam-splitter outputs, the joint spin--photon state can be written as
\[
\frac{1}{2}\!\left(
\ket{\psi_{00}}\ket{00}
+ \ket{\psi_{01}}\ket{01}
+ \ket{\psi_{10}}\ket{10}
+ \ket{\psi_{2}}\ket{\mathrm{noon}}
\right),
\]
where
\begin{align*}
\begin{split} &\ket{\psi_{00}} = \ket{\downarrow\downarrow}, \quad \ket{\psi_{01}} = \frac{\ket{\uparrow\downarrow} + \ket{\downarrow\uparrow}}{\sqrt 2},\\ &\ket{\psi_{10}} = \frac{\ket{\uparrow\downarrow} - \ket{\downarrow\uparrow}}{\sqrt 2}, \quad \ket{\psi_{2}} = \ket{\uparrow\uparrow}. \end{split}
\end{align*}
where $\ket{\mathrm{noon}} = \frac1{\sqrt 2} (\ket{20}+\ket{02})$. Heralding on a single-photon detection event (outcomes $01$ or $10$) projects the remote spins onto a maximally entangled Bell state, thereby generating long-distance entanglement without requiring direct interaction between the qubits. 

\paragraph{Zero-order analysis} In a general setting, where the beam splitter angle is $\theta$, the successful events (i.e., measurements $\mathbf m \in \{(1,0) , (0,1)\}$, are associated with the following $\mathbf D$ matrices:
\begin{align}
\begin{split}
D_{(0,1)} = \cos\theta \sigma_{\uparrow,1} + \sin\theta \sigma_{\uparrow, 2},\\ D_{(1,0)} = -\sin\theta \sigma_{\uparrow,1} + \cos\theta \sigma_{\uparrow, 2},
\end{split}
\end{align}
where $\sigma_{\uparrow, i} := \ket{\uparrow}\bra{\uparrow'}_i$ corresponds to the lowering operators of the $i$-th emitter. Here, and also in the next section, we go to the rotating frame given by the system Hamiltonian, so that 
\begin{align}
H_{\mathrm{eff}} = -\frac{\gamma}{2} (\ket{\uparrow'}\bra{\uparrow'}_1 + \ket{\uparrow'}\bra{\uparrow'}_2).
\end{align}
Furthermore, following \cite{wein2020analyzing}, we wait for a long time after the detection, such that the system is no longer in the excited state. This is equivalent to applying the \textit{long waiting} channel $\mathcal E_{\mathrm{LW}}$ defined as
\begin{align}
\mathcal E_{\mathrm{LW}}[\bullet] = \Pi_{\uparrow'}^\perp \bullet \Pi_{\uparrow'}^\perp + \bra{\uparrow'}\bullet\ket{\uparrow'} \ket{\uparrow}\bra{\uparrow}
\end{align}
which ensures that all the population from $\ket{\uparrow'}$ is brought to $\ket{\uparrow}$. Note that since we have $H_{\mathrm{eff}}$ and $\mathbf D$, we can compute the conditional states via \cref{eq:zero-order-closed-form}. For simplicity, we focus only on $\mathbf m = (0,1)$ measurement result. We consider the initial state (i.e., the state right after the $\pi$-pulse)
\begin{align}
\rho_{\mathrm{init}} = (\cos\vartheta \ket{\downarrow} + \sin\vartheta \ket{\uparrow'})^{\otimes 2} (\mathrm{h.c.}).
\end{align}
Following our zero-order solutions in \cref{sec:zero-order}, we obtain that, in the zero-th order, the state after $\mathcal E_{\mathrm{LW}}$ is given by
\begin{align}
\begin{split}
\rho^{(0)}_{\mathbf m = (0,1)} = &\rho^{(0)}_{\uparrow\uparrow} \ket{\uparrow \uparrow}\bra{\uparrow\uparrow} + \rho_{\uparrow\downarrow}^{(0)}\ket{\uparrow\downarrow}\bra{\uparrow\downarrow}\\
&+ \rho^{(0)}_{\downarrow\uparrow}\ket{\downarrow\uparrow}\bra{\downarrow\uparrow} + (\rho_c^{(0)} \ket{\downarrow\uparrow}\bra{\uparrow\downarrow} + \mathrm{h.c.}),
\end{split}
\end{align}
where
\begin{align}
\begin{split}
\rho^{(0)}_{\uparrow\uparrow} &= (1-e^{-\gamma T_d}) e^{-\gamma T_d} \sin^4\vartheta \\
\rho^{(0)}_{\uparrow\downarrow} &= (1-e^{-\gamma T_d}) \sin^2\vartheta\cos^2\vartheta \cos^2\theta \\
\rho^{(0)}_{\downarrow\uparrow} &= (1-e^{-\gamma T_d}) \sin^2\vartheta\cos^2\vartheta \sin^2\theta\\
\rho_c^{(0)} &= (1-e^{-\gamma T_d}) \sin^2\vartheta \cos^2\vartheta \sin\theta \cos\theta.
\end{split}
\end{align}
In what follows, we consider a particular error model and show that the resulting density matrix matches the previously known results from the literature, up to the first order.

\paragraph{First-order analysis} We consider a pure optical dephasing error with rate $\gamma^\ast$. Such an error is described by the Lindbladian term
\begin{align}
2\gamma^\ast \left( \mathcal D(\ket{\uparrow'}\bra{\uparrow'}_1) + \mathcal D(\ket{\uparrow'}\bra{\uparrow'}_2)  \right).
\end{align}
Substituting into the first order evolution equation \cref{eq:first-order-solution}, and applying $\mathcal E_{\mathrm{LW}}$ gives
\begin{align}
\rho_{\mathbf m = (0,1)} = \rho_{\mathbf m=(0,1)}^{(0)} + \gamma^\ast \rho_{\mathbf m = (0,1)}^{(1)} + O(\gamma^\ast{}^2),
\end{align}
where
\begin{align}\label{eq:our-fo}
\begin{split}
\rho_{\mathbf m = (0,1)}^{(1)} = &\frac{2}{\gamma} \sin^2\vartheta \cos^2\vartheta \sin\theta \cos\theta \\
&\qquad(-(1-e^{-\gamma t}) + \gamma t e^{-\gamma t}) \ket{\uparrow\downarrow}\bra{\downarrow\uparrow} \\
&+ \mathrm{h.c.}.
\end{split}
\end{align}
Next, we show that our solution is in agreement with the result of \cite{wein2020analyzing}, up to the first order.

\paragraph{First-order comparison with \cite{wein2020analyzing}}Now, let us turn to the solution in \cite{wein2020analyzing} and compute the error for $\gamma^\ast$. In that solution we have
\begin{align}
\tilde C(T_d) = \frac{2\gamma}{2\gamma + 2\gamma^\ast} (1-e^{-T_d(\gamma + \gamma^\ast)}) \approx (1-\frac{\gamma^\ast}{\gamma}) (1-e^{-T_d\gamma}(1-T_d \gamma^\ast)) =  (1-e^{-T_d\gamma}) + \frac{\gamma^\ast}{\gamma} ( -(1-e^{-\gamma T_d}) + \gamma T_d e^{-T_d \gamma}),
\end{align}
where we have used $\frac{\gamma}{\gamma + \gamma^\ast} = (1+\frac{\gamma^\ast}{\gamma})^{-1}$ along with the approximation $(1+x)^n \approx 1+nx$ for small $|x|$. Also, $e^x \approx 1+x$ for small $|x|$. Therefore, the first-order error, according to \cite{wein2020analyzing} is
\begin{align}
\gamma^\ast ( -(1-e^{-\gamma T_d}) + \gamma T_d e^{-T_d \gamma}) \cdot \frac18\sin^2(2\vartheta) \sin(2\theta) \ket{\uparrow\downarrow}\bra{\downarrow\uparrow} + h.c.
\end{align}
which matches our expression \cref{eq:our-fo}.

\section{Phase shifter error analysis in a RUS gate}\label{sec:phase-shifter}

We show how one can readily compute the effects of phase shifter error in a RUS gate. This is the error in tuning the parameter $\phi$ in \cref{eq:U-of-rus}. This error affects only the third and fourth columns of $U$ in \cref{eq:U-of-rus}, meaning that only $D_3$ and $D_4$ get affected. Also, for simplicity, we assume all other sources of error are negligible, and hence, we can use the zero-order solutions obtained in \cref{eq:zero-order-closed-form} (as the phase shifter error can be absorbed into perturbation on $\mathbf D$ operators, and there is no added Lindbladian to the system). Since only $D_3, D_4$ are perturbed, all patterns that detect photons in the first two detectors only, remain unaffected by this error. This can be also viewed intuitively from \cref{fig:RUS-levels-and-table}, as the phase shifter element can only cause errors on the errors on the top two modes. Let us write $\phi = \frac{\pi}{2} + \delta$, and write $D_3, D_4$ as
\begin{align}
\begin{split}
D_3 &= \frac{1}{2} (\sigma_{r,1} - \sigma_{\ell,1} + i e^{i\delta} \sigma_{r,2} - ie^{i\delta} \sigma_{\ell,2}),\\
D_4 &= \frac{1}{2} (\sigma_{r,1} - \sigma_{\ell,1} - i e^{i\delta} \sigma_{r,2} + ie^{i\delta} \sigma_{\ell,2}).
\end{split}
\end{align}
Using \cref{eq:zero-order-closed-form}, we can write the conditional map
\begin{align}
\mathcal P_{(1010)}(t,0)[\bullet] = f_{(1010)}(t) \exp(-iH_{\mathrm{eff}}t) D_1 D_3 \bullet D_1^\dag D_3^\dag \exp(iH_{\mathrm{eff}}t).
\end{align}
It is straightforward to verify that $\exp(-iH_{\mathrm{eff}}t) D_1 D_3 = D_1 D_3$, and hence, $\mathcal P_{(1010)}(t,0)[\bullet] = f_{(1010)}(t) D_1 D_3 \bullet D_1^\dag D_3^\dag$. Letting $\mathcal P_{(1010)}^{(0)}$ denote the channel in the ideal case (i.e., when $\delta=0$), we write the error channel as $\mathcal E_{\mathrm{err}} = \mathcal P_{(1010)}^{(0)} {}^{-1}\circ \mathcal P_{(1010)}$, and then by calculating the Choi matrix (say $J$) and looking at elements such as $\bbra{P} J \kket{P}$ we obtain Pauli error $P$ associated with our phase error $\delta$. Doing so, we obtain that all Pauli error weights are zero, except for the $ZZ$ error, which admits a simple form
\begin{align}
\chi_{ZZ} = \sin^2\left(\frac{\delta}{2}\right).
\end{align}
There are also non-diagonal entries of $II,ZZ$ and $ZZ,II$ that we compute:
\begin{align}
\chi_{ZZ,II} = \chi_{II,ZZ} = -\frac12 \sin\delta.
\end{align}

We highlight that the $(2000)$ pattern is not affected by the phase shifter error, and hence, there would be no Pauli error for that conditional map.

Lastly, we highlight that our approach can be used to derive analytical expressions for the combined effects of coherent and incoherent errors. For instance, one can consider the effects of phase shifter imperfections together with coherent errors studied in the main text.

\end{document}